\documentclass[12pt]{article}
\usepackage{newtxtext,newtxmath}

\usepackage{graphicx}
\usepackage[ruled, lined, linesnumbered]{algorithm2e}
\usepackage{bm}

\usepackage[letterpaper,margin=1in]{geometry}

\renewenvironment{abstract}
	{\quotation}
	{\endquotation}

\date{}

\makeatletter
\renewcommand{\fnum@figure}{\textbf{Figure \thefigure}}
\renewcommand{\fnum@table}{\textbf{Table \thetable}}
\makeatother

\usepackage{scicite}

\usepackage{url}
\usepackage{caption}
\usepackage{float}
\usepackage{placeins}

\usepackage{fancyhdr}
\def\scititle{
     Rescheduled, not redefined: The moving plateau of old-age mortality
}
\title{\bfseries \boldmath \scititle}

\author{
	Silvio~C.~Patricio$^{1\ast}$ and Trifon~I.~Missov$^{1}$\and
	\small$^{1}$Interdisciplinary Center on Population Dynamics, University of Southern Denmark, Odense, 5230, Denmark.\and
	\small$^\ast$Corresponding author. Email: silca@sam.sdu.dk\and
	\small$^\dagger$These authors contributed equally to this work.
}

\begin{document} 

\maketitle
\begin{abstract} \bfseries \boldmath \noindent
    Whether the risk of death keeps climbing at extreme ages or levels off has divided researchers for a century. We show this conflict reflects a moving target. Using cohort data from twelve low-mortality populations, we find that mortality deceleration and plateau onset shift steadily later across cohorts born from the mid-19th to the early-20th century. The data support a plateau across the full cohort range, though evidence weakens for the youngest, incompletely observed cohorts. In these younger cohorts, deceleration begins near age 100, and the fitted plateau begins beyond age 108. Because past studies focused on different cohorts and fixed age ranges, they sampled different phases of the exact same shift. The mortality plateau has no fixed age: it has been rescheduled.
\end{abstract}



\noindent

At extreme ages, does the risk of death keep climbing, or does it level off? This is one of the oldest open questions in human longevity, and it is becoming more pressing, as the number of people reaching age~100 is projected to increase sixfold over the next three decades \cite{un_wpp2024}. This rise reflects both the number who survive to old age and a late-life mortality schedule that has moved to later ages \cite{canudas2008modal}. The century-old disagreement over whether mortality levels off may therefore turn less on what each study found than on when it looked.

For decades, studies of mortality at extreme ages have reached opposing conclusions about whether death rates keep rising, decelerate, or level off. Many report a deceleration \cite{kannisto1994development, horiuchi1998deceleration, steinsaltz2006understanding, vaupel1998biodemographic}, and several find that rates settle onto a plateau \cite{vaupel1998biodemographic, gampe2010human, gampe2021mortality, rootzen2017human, barbi2018plateau, belzile2022there}, with support from high-quality data on Italian longevity pioneers \cite{barbi2018plateau} and on several European and North American populations \cite{alvarez2021regularities}.

Others reject deceleration altogether \cite{gavrilova2015biodemography, gavrilov2011mortality, gavrilov2019new}. They attribute the apparent flattening to age-reporting errors and the small number of deaths at the oldest ages \cite{gavrilov2019late, newman2018plane, newman2018errors}; a recent analysis of French cohorts likewise finds no evidence for a plateau \cite{Danetal23}. Each study, however, observes a narrow window: a limited set of cohorts over a limited age range. These windows rarely overlap.

A natural explanation is that late-life mortality itself moves. The full schedule---from deceleration to plateau onset---may shift to older ages as deaths are postponed and longevity rises \cite{horiuchi2013modal, canudas2008modal, vazquez2024longevity, patricio2024makeham}. Lynch and Brown showed that the timing of mortality slowing is not fixed \cite{lynch2001reconsidering}; later studies documented variation across cohorts and by sex \cite{li2013mortality, wrigleyfield2014mortality}. Whether deceleration and plateau onset move together across birth cohorts and populations remains unresolved. Studies anchored to fixed cohorts and ages may therefore observe different phases of the same schedule---one while mortality is still slowing, another after it has leveled off.

We test this hypothesis with cohort mortality from twelve low-mortality populations and cohorts born from the mid-19th to the early-20th century. We model each cohort with the gamma--Gompertz model \cite{vaupel1979impact}, in which selection produces deceleration and a plateau: frailer individuals die earlier, the surviving population becomes less frail, and population mortality levels off even as individual risk keeps rising. A Bayesian state-space model allows the parameters to follow a random walk across cohorts \cite{patricio2026rhythm, augermethe2021guide, haakman2025frequency}. This structure shares information across adjacent cohorts while preserving uncertainty where deaths become sparse above age~100.

For each cohort we estimate three landmarks: the age at which mortality begins to decelerate, the age at which it approaches a plateau, and the level of that plateau. One parameter, the variance of frailty, determines whether a plateau exists. Setting this variance to zero reduces the model to a pure Gompertz law \cite{gompertz1825xxiv}, whose tail rises without limit. This nesting lets the data evaluate a plateau in each cohort rather than building it into the analysis.

\section*{Data and Methods}
We analyze cohort mortality from twelve low-mortality populations---Australia, Canada, Denmark, England and Wales, France, Italy, Japan, the Netherlands, New Zealand, Norway, Sweden, and Switzerland---using the Human Mortality Database \cite{hmd}. For each population and sex, we use age-specific death counts and exposures from age~65 onward. Beginning at age~65 places the lower bound before the modal age at death while excluding younger ages, where nonsenescent risks contribute more strongly to mortality \cite{patricio2026rhythm}. The first included cohort varies by population, from 1840 to 1868; every series ends with the 1910 cohort (table~\ref{tab:cohorts}).

We fit the gamma--Gompertz model \cite{vaupel1979impact} to each of the twenty-four population--sex series. Three parameters describe each cohort: the baseline mortality level, the Gompertz slope, and the variance of frailty. From them we derive the ages of deceleration and plateau onset and the plateau level.

Within each series, these parameters follow a first-order random walk on the log scale \cite{patricio2026rhythm}. Linking adjacent cohorts allows well-observed cohorts to inform those with few deaths at the oldest ages. We fit the model with Hamiltonian Monte Carlo in Stan \cite{stan}. The Supplementary Materials reports the priors, sampling procedure, convergence checks, and fit diagnostics (figs.~\ref{fig:fit_australia}--\ref{fig:fit_Switzerland}).

We test for a plateau with a spike-and-slab prior on the frailty variance \cite{mitchell1988bayesian, george1993variable}. The spike fixes the variance at zero, producing a pure Gompertz tail with no plateau; the slab keeps the variance positive and follows the same cohort-to-cohort random walk as the continuous model. A shared mixture weight with a uniform prior gives the two states equal prior weight on average. For each cohort, we estimate the posterior probability of the no-plateau state (fig.~\ref{fig:spike_slab}).

\section*{Results}\label{results}
Across all populations, late-life mortality begins to slow and approaches a plateau at progressively older ages in cohorts born between the mid-19th and the early-20th century. The plateau level remains broadly stable in most populations but rises in the youngest ones. We first test whether the data support a plateau, then examine changes in its timing and level.

The spike-and-slab model compares positive frailty variance, which produces a plateau, with zero variance, which makes the population homogeneous with respect to frailty and yields a pure Gompertz trajectory. For every cohort, the posterior probability of the no-plateau state lies far below its prior probability. The data therefore support positive frailty variance, mortality deceleration, and a late-life plateau (fig.~\ref{fig:spike_slab}).

The no-plateau probability rises in some of the youngest cohorts but remains below the prior benchmark. These cohorts often contribute more deaths at ages already reached, but they remain incompletely observed at the highest ages, where a plateau is most clearly distinguished from continued Gompertz growth (fig.~\ref{fig:counts}). Support for a plateau thus extends across the cohort range, although it weakens for the most recent cohorts.

\subsubsection*{The late-life schedule shifts to later ages}

\begin{figure}[!htb]
    \centering
    \includegraphics[width=\textwidth]{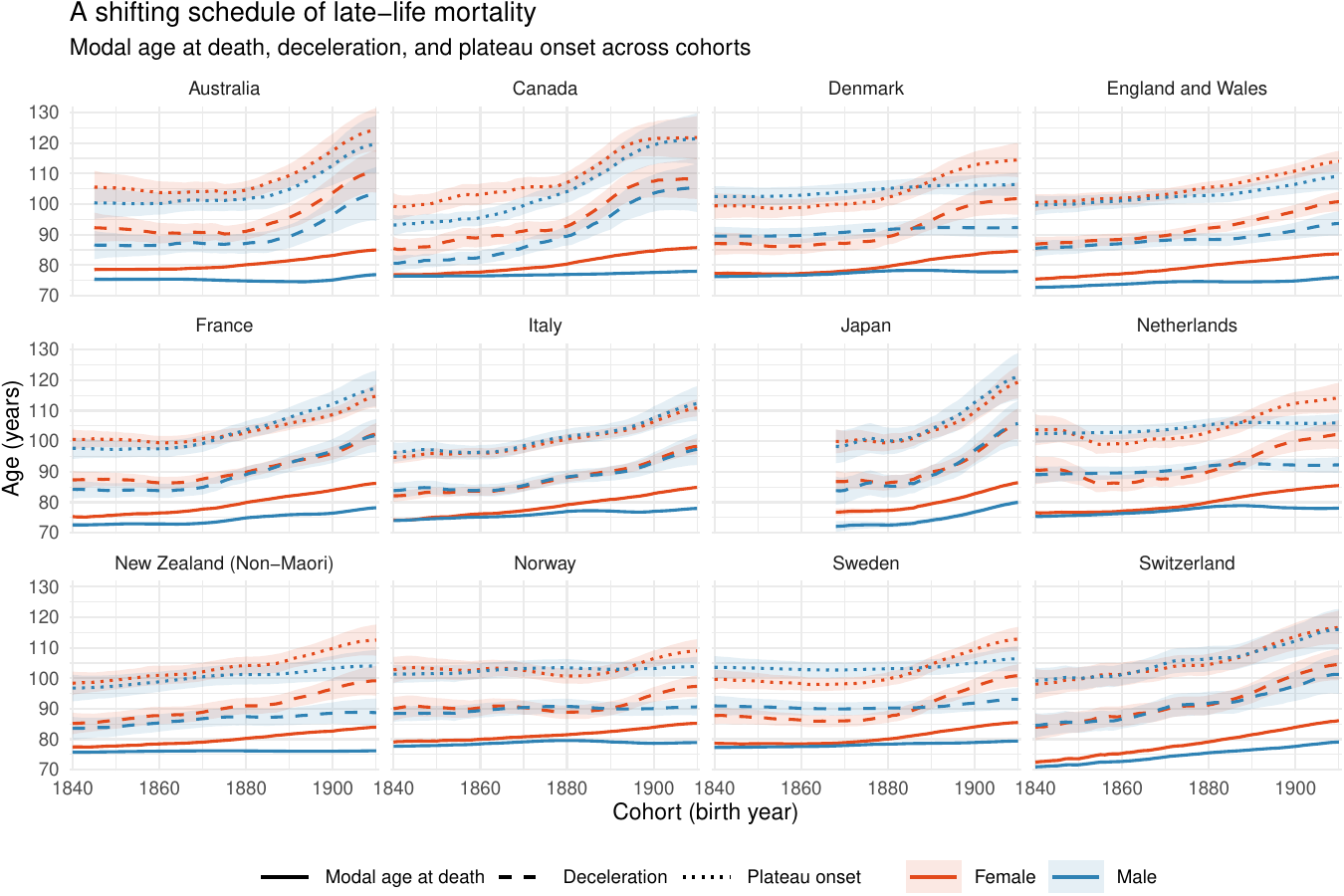}
    \caption{\textbf{A shifting schedule of late-life mortality across cohorts.} Trajectories show the modal age at death (solid), age of mortality deceleration (dashed), and plateau onset (dotted) in twelve populations. Cohorts span 1840--1910, except in Australia (1845--1910) and Japan (1868--1910). Female trajectories are red and male trajectories blue; shaded bands give 95\% credible intervals.}
    \label{fig:schedule}
\end{figure}

Figure~\ref{fig:schedule} traces three landmarks for each population and sex: the modal age at death, the age of mortality deceleration, and the plateau onset. Across every population and both sexes, the last two move to later ages across cohorts. The modal age also rises, although it remains nearly flat for men in many populations. The landmarks retain their order---mode, deceleration, plateau---so the schedule shifts later without rearranging its parts.

\begin{figure}[!b]
\centering
 \includegraphics[width=\textwidth]{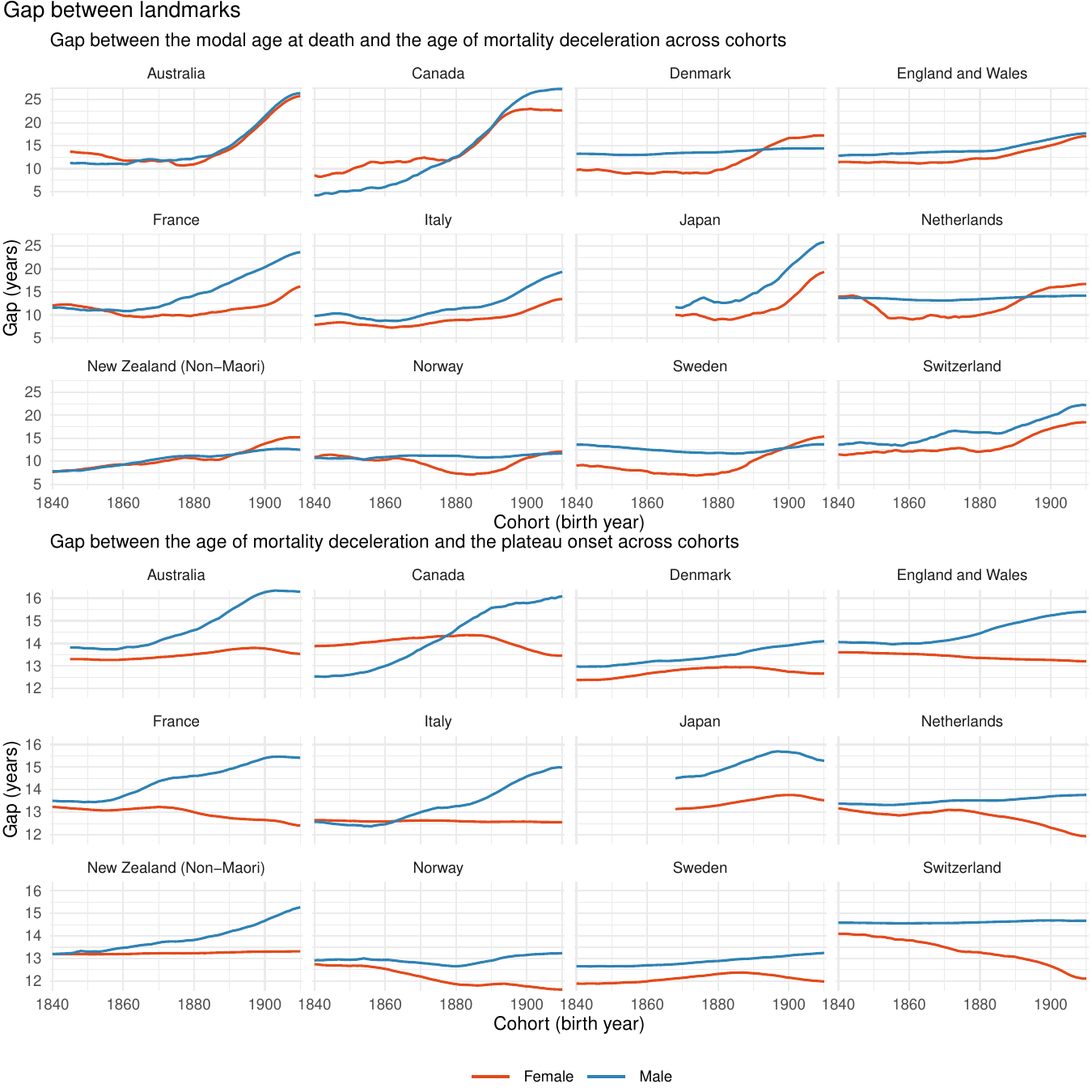}
  \caption{\textbf{Intervals between landmarks of the late-life schedule.} The top panels show the interval from the modal age at death to mortality deceleration; the bottom panels show the interval from deceleration to plateau onset. Cohorts span 1840--1910, except in Australia (1845--1910) and Japan (1868--1910), across twelve populations. Female trajectories are red and male trajectories blue. The mode-to-deceleration interval widens in every population, showing that the extreme-age tail shifts later more rapidly than the center of the death distribution. The deceleration-to-plateau interval widens for men in most populations but remains stable or narrows for women, with the clearest contrasts in Australia, Canada, France, Italy, Japan, and Switzerland.}
 \label{fig:gaps}
\end{figure}

The landmarks move at different rates. Averaged across cohorts, the modal age rises by about 0.09 years per cohort, compared with roughly 0.23 years for deceleration and 0.25 years for plateau onset (table~\ref{tab:mean_steps}). The extreme-age landmarks therefore pull away from the mode: survival extends faster in the tail than near the age at which most deaths occur.

Women show larger gains in the modal age. For men, however, a nearly flat mode and a rising deceleration age widen the gap between the center and tail by at least as much as for women, and by more in some populations (Figs.~\ref{fig:schedule} and~\ref{fig:gaps}, top panel). The deceleration and plateau ages rise at least as fast for men as for women in several populations, including Canada, France, and Japan (table~\ref{tab:mean_steps}).

The interval between deceleration and plateau onset follows a distinct, sex-specific pattern (Fig.~\ref{fig:gaps}, bottom). It widens for men in most populations because plateau onset shifts later more rapidly than deceleration. For women, it remains stable or narrows. The contrast is strongest in Australia, Canada, France, Italy, Japan, and Switzerland.

Observed deaths anchor the shift across most of the cohort range. For most cohorts, plateau onset lies at or below the highest age with recorded deaths (fig.~\ref{fig:counts}); deceleration occurs earlier still, below age~100 except in the most recent cohorts. The central finding that the schedule moves later therefore does not depend on extrapolation at the sparsest ages.

Only in the youngest cohorts does plateau onset extend beyond the observed frontier. Their fitted schedules begin to decelerate near age~100 and reach the plateau beyond age~108; in a few populations, the estimated onset exceeds age~115. Deaths are sparse at these ages, and individuals from the youngest cohorts may still be alive, leaving their extreme-age mortality only partly observed (fig.~\ref{fig:counts}). The resulting credible intervals are wide, so these onsets are projections of the fitted hazard rather than settled values. Short cohort series in Japan and Australia and sparse counts in New Zealand add uncertainty. The shift itself, however, is established before the schedule reaches this frontier.

\subsubsection*{The level of the plateau}
\begin{figure}[!tb]
    \centering
    \includegraphics[width=\textwidth]{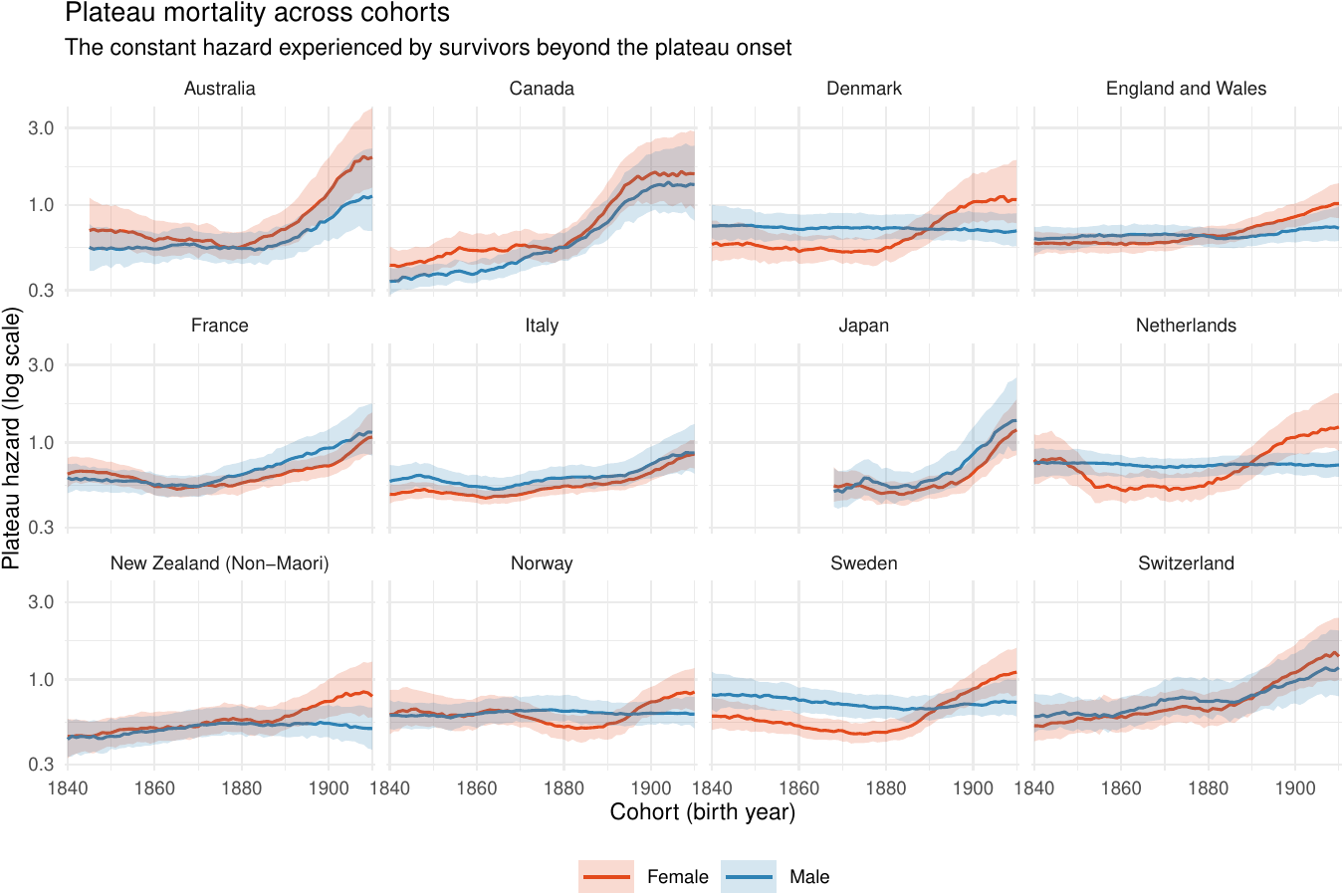}
    \caption{\textbf{Mortality plateau level across cohorts.} Values equal the inverse of remaining life expectancy at plateau onset, calculated from the fitted mortality tail (Supplementary Materials). Cohorts span 1840--1910, except in Australia (1845--1910) and Japan (1868--1910), across twelve populations. Values are shown on a logarithmic scale. Female trajectories are red and male trajectories blue; shaded bands give 95\% credible intervals.}
    \label{fig:level}
\end{figure}

Figure~\ref{fig:level} summarizes the plateau level for each cohort with a single constant hazard representing mortality across the fitted tail beyond plateau onset (see the supplementary material). This quantity describes the tail as a whole, not the instantaneous hazard at the onset age. A value near 0.7 corresponds to an annual survival probability of about one-half; higher values imply lower annual survival.

In Denmark, Sweden, Norway, the Netherlands, and England and Wales, the plateau level stays stable across most cohorts and rises only among the youngest. Women drive most of this late increase; male levels remain nearly flat. In Australia, Canada, France, Italy, Japan, and Switzerland, the increase begins earlier and appears in both sexes, reaching about twice the mid-19th-century level in some cohorts. Wide intervals and sparse deaths prevent a clear classification of New Zealand.

Timing and level move together. The plateau level rises in the populations where the deceleration-to-plateau interval diverges by sex and remains stable where the interval does not diverge.

The steepest increases occur in the youngest cohorts, whose estimated plateaus begin beyond age~108 and rely on sparse observations (fig.~\ref{fig:counts}). Wide credible intervals show that part of this rise may reflect limited information rather than a change in mortality.

Even with this uncertainty, posterior estimates for recent cohorts in these populations exceed the level near~0.7 reported in validated studies of longevity pioneers \cite{belzile2021human,barbi2018plateau, rootzen2017human, belzile2022there}. Earlier cohorts remain near the reported 0.6--0.8 range, whereas recent estimates imply annual survival falling from roughly one-half toward one-third. The plateau level therefore varies by cohort rather than taking a single value.

\section*{Discussion}\label{discussion}

Treating plateau onset as cohort-specific resolves much, but not all, of the century-long dispute over extreme-age mortality. Studies that analyze fixed ages can reach different conclusions because they observe different phases of a moving schedule.

Our results offer a direct reinterpretation of the French cohorts studied by Dang et al.\ \cite{Danetal23}. Their threshold at age~105 may sample the deceleration phase after plateau onset has shifted to older ages. Their fitted Gompertz slope, about 0.06, lies below the value near 0.1 typical of adult mortality \cite{patricio2026rhythm} and is consistent with the deceleration they report. Their result therefore need not imply that the plateau is absent; an analysis anchored at older ages might have yielded evidence for it.

Barbi et al.\ found a plateau from age~105 and a small downward cohort trend in its level \cite{barbi2018plateau}. In our results that trend is read as the signature of a plateau onset moving to later ages, not a plateau whose level is falling.

Building on Lynch and Brown's U.S. period analysis, we trace the late-life mortality schedule across cohorts in twelve populations \cite{lynch2001reconsidering}. Our framework distinguishes mortality deceleration from the later approach to the plateau and relates both to the modal age at death. This extends earlier work from the timing of a single turning point to the full sequence from the modal age through deceleration to plateau onset.

Yet differences in cohort and age coverage cannot explain every disagreement. Studies using the same data can still reach different conclusions because of age validation or model choice; the debate over the Italian plateau is one example \cite{barbi2018plateau, newman2018errors}. Our analysis does not settle that dispute.

The spike-and-slab analysis addresses one part of the model-choice concern. A gamma--Gompertz hazard levels off whenever frailty variance is positive, but our test also allows for zero variance and a pure Gompertz tail. Across every cohort, the posterior probability of that no-plateau state remains far below its prior value, although the contrast weakens in the youngest cohorts (fig.~\ref{fig:spike_slab}). The functional form alone therefore does not account for the evidence of leveling.

The comparison therefore focuses on the two patterns supported by empirical studies of human mortality: continued Gompertz growth and deceleration toward a plateau. We do not model a declining tail, for which current human data provide no clear evidence (Supplementary Materials).

\subsubsection*{The schedule shifts later, but not uniformly}

In Denmark, Sweden, Norway, the Netherlands, and England and Wales, deceleration and plateau onset move later together, leaving the interval between them nearly unchanged. Within the model, this interval varies inversely with the Gompertz slope (Eq.~\eqref{eq:spacing}); its stability therefore indicates little change in that slope. In these populations, the schedule shifts later with little change in its late-life shape.

Postponement, however, is not uniform across the mortality schedule. The shifting-mortality hypothesis holds that as deaths move later, the distribution retains its shape around the modal age, keeping its spread and remaining life expectancy at the mode broadly stable \cite{canudas2008modal, vazquez2024longevity}. Our extreme-age landmarks rise faster than the mode (table~\ref{tab:mean_steps}). 

The distribution therefore does not shift rigidly across all ages: its central mass can retain a stable width while mortality in the far-right tail shifts later relative to the mode. This pattern does not imply a change in the maximum human lifespan, which our analysis does not estimate. In demographic terms, it represents local decompression above the mode. The gap between typical death and extreme-age mortality widens, so the modal age and measures of compression around it may understate the pace of mortality postponement among the oldest survivors.

\subsubsection*{Sex divergence and historical shocks}
A second pattern appears in Australia, Canada, France, Italy, Japan, and Switzerland: the interval between deceleration and plateau onset diverges by sex, and the plateau level rises (Figs.~\ref{fig:gaps} and~\ref{fig:level}). Within the model, a wider interval is equivalent to a lower fitted Gompertz slope; it is not an independent measure of selection duration. New Zealand fits neither pattern because sparse deaths leave its estimates too uncertain.

The timing of this divergence overlaps the two World Wars and the 1918 influenza pandemic, making historical shocks a plausible explanation rather than an identified cause. Such events affect several neighboring cohorts at different ages, so a cohort series may record them as gradual departures from trend rather than isolated jumps \cite{patricio2026rhythm}. 

Historical shocks could contribute to the sex divergence through differences in exposure, mortality selection, and lasting harm. Evidence from the 1918 pandemic shows that mortality selection affected men and women differently and helped shape later differences in their mortality \cite{noymer2000influenza}. More generally, severe shocks can change the composition of surviving cohorts through selective mortality \cite{zarulli2012mortality, zarulli2013effect}, whereas early infectious and inflammatory exposure may have lasting effects on old-age mortality \cite{finch2004inflammatory}. 

These pathways are plausible, but neither maps uniquely onto the fitted Gompertz slope or plateau level. Our aggregate model cannot distinguish them. The observed rise in the plateau level is compatible with lasting damage, but it does not identify that mechanism. Switzerland also diverges despite remaining neutral in both wars, so combat exposure cannot explain the full pattern.

France and Sweden provide a suggestive contrast: the interval widens across the affected French cohorts but remains nearly flat in Sweden, which stayed out of both wars (Fig.~\ref{fig:gaps}). This pattern is consistent with an effect of historical shocks but does not establish one.

\subsubsection*{When, not how many}

Our results separate the timing of extreme-age mortality from the number of people who reach it. While deceleration remains within the well-observed range, the share of each cohort surviving from age~65 to its own deceleration age stays broadly stable (fig.~\ref{fig:sirvival}). This share declines in the youngest cohorts, but their deceleration ages extend into sparsely observed ages (fig.~\ref{fig:counts}), so incomplete follow-up may drive the decline. A later schedule can therefore move more survivors past a fixed age such as~100 even when the fraction reaching each cohort's own deceleration frontier changes little.

Population aging depends on both timing and volume. Our analysis identifies timing: holding the number of survivors fixed, mortality deceleration and the plateau occur at later ages in successive cohorts. It does not measure the timing of disability, care needs, or fiscal costs. Nor does it determine how many people reach extreme age. That number also depends on birth-cohort size and survival earlier in life, and it can rise even when the share reaching each cohort's own late-life frontier remains stable.

The projected sixfold rise in centenarians \cite{un_wpp2024} can therefore combine larger cohorts, improved survival to old age, and a mortality schedule shifted beyond age~100. Our findings do not show whether the need for care and support will rise, fall, or remain stable. They show that counts and timing describe distinct dimensions of population aging and must be assessed together.

\subsubsection*{Interpreting later timing}

Women lead the shift in the modal age, and because more women reach extreme ages than men, the female tail is better observed. The extreme-age landmarks are not led by women, and for several populations they rise as fast or faster for men (table~\ref{tab:mean_steps}).

A later schedule is not, on its own, necessarily a healthier one. The landmarks describe when late-life mortality changes, not the health of those who reach these ages. If later meant better, the Nordic countries, which reach these landmarks earliest, would rank among the worst; instead they are among the healthiest of these cohorts \cite{jasilionis2024increases, popham2013health}, with the highest survival to old age. These estimates therefore describe timing, not a ranking of population.

\subsubsection*{Limits and complementary evidence}
Observed deaths anchor our estimates, but most populations record fewer than ten deaths per age--cohort cell above about age~108 (fig.~\ref{fig:counts}). Plateau onsets beyond the observed range are therefore projections of the fitted hazard. The death records behind these counts are administrative registers of uneven quality across countries and cohorts, and are not individually validated at the oldest ages, so the sparsest cohorts, the most recent in Japan and New Zealand above all, consequently carry the widest uncertainty. Future work should combine these aggregate series with individually validated records of semi-supercentenarians and supercentenarians.

Part of the disagreement is one of method and data, not only of where the threshold is set. Studies of validated individual records---including IDL supercentenarians, Italian semi-supercentenarians, and French cohorts---typically begin at a fixed age and pool cohorts or allow a single cohort trend \cite{barbi2018plateau, belzile2021human, Danetal23, belzile2022there, rootzen2017human}. Such designs estimate mortality at the phase captured by that threshold but cannot directly track a moving landmark. Our state-space model instead follows the schedule across cohorts and lets adjacent cohorts inform sparse ages.

The approaches are complementary. Our model traces how the plateau moves across cohorts but cannot validate individual deaths at extreme ages. Validated records establish mortality where ages and identities have been checked but provide less leverage for estimating these cohort trends. They also provide an external check on our extrapolations: our recent-cohort onsets fall near the ages beyond~108 at which validated data find mortality becoming approximately constant \cite{belzile2021human}. The agreement is informative, but the validated records do not enter our model and therefore do not remove its uncertainty.

\section*{Conclusion}
Late-life mortality is not fixed at particular ages. Across cohorts, both deceleration and plateau onset shift to later ages, in some populations by nearly a decade. This postponement helps explain why studies of different cohorts and age ranges reached opposing conclusions: they observed different phases of the same schedule. The evidence supports a plateau, but its onset depends on cohort.

Most populations preserve the late-life shape as the schedule moves; others show sex-specific changes in the fitted Gompertz slope and plateau level. These changes coincide with major historical shocks, although the data do not identify their causes. The extreme-age tail also shifts faster than the modal age, widening the distance between the survival frontier and the center of the death distribution. As centenarian populations grow, this distinction between a moving schedule and a changing population size becomes increasingly important.

Extreme old age has not been redefined. It has been rescheduled.


\clearpage 

%
\bibliography{science_template} 
\bibliographystyle{sciencemag}

%
%
%
%
%
%


\section*{Acknowledgments}
\paragraph*{Funding:}
This research was supported by the AXA Research Fund through the AXA Chair in Longevity Research and by the SCOR Foundation for Science through the SCOR Chair in Mortality Research. The article is also funded by the European Union (ERC, Born Once – Die Once, Grant agreement ID 101043983). The views and opinions expressed are solely those of the authors and do not necessarily reflect those of the European Union or the European Research Council Executive Agency. Neither the European Union nor the granting authority can be held responsible for them.

\paragraph*{Author contributions:}
Conceptualization, Methodology and Visualization: S.C.P.; 
Project administration and Supervision: T.I.M.; 
Writing – original draft, review and editing: S.C.P. and T.I.M.

\paragraph*{Competing interests:} There are no competing interests to declare.

\paragraph*{Data and materials availability:}
The data come from the publicly available Human Mortality Database
(\url{https://www.mortality.org}). The code used to prepare the data, fit the model, and generate the figures and tables, is available on GitHub
(\url{https://github.com/scpatricio/dynamic_shifts_in_mortality_patterns}).


\subsection*{Supplementary Materials}
Materials and Methods\\
Supplementary Figures\\
Supplementary Tables\\


\newpage


\renewcommand{\thefigure}{S\arabic{figure}}
\renewcommand{\thetable}{S\arabic{table}}
\renewcommand{\theequation}{S\arabic{equation}}
\renewcommand{\thepage}{S\arabic{page}}
\setcounter{figure}{0}
\setcounter{table}{0}
\setcounter{equation}{0}
\setcounter{page}{1} 


\begin{center}
\section*{Supplementary Materials for\\ \scititle}

	Silvio~C.~Patricio$^{1\ast}$ and 
    Trifon~I.~Missov$^{1}$\\
	\small$^{1}$Interdisciplinary Center on Population Dynamics, University of Southern Denmark, Odense, 5230, Denmark. \\
	\small$^\ast$Corresponding author. Email: silca@sam.sdu.dk \hspace{5mm} 
	\small$^\dagger$These authors contributed equally to this work.
\end{center}

\subsubsection*{This PDF file includes:}
Material and Methods\\
Supplementary Figures\\
Supplementary Tables\\


\newpage

\section*{Material and Methods}

\subsection*{The mortality model}
We represent senescent mortality with the gamma--Gompertz model \cite{vaupel1979impact}. Individuals share a Gompertz age pattern but differ in fixed, unobserved frailty. An individual with frailty $Z=z$ has hazard
\begin{equation}
    \mu(x\mid z)=z\,a\,e^{\,bx},
    \label{eq:indiv}
\end{equation}
where $a>0$ is the baseline hazard and $b>0$ is the Gompertz slope, which controls how fast individual risk rises with age. We assume that frailty is fixed at birth and follows a gamma distribution with $\mathbb{E}(Z)=1$ and $\mathbb{VAR}(Z)=\gamma$. The variance $\gamma$ measures heterogeneity in susceptibility to death.

Because frailer individuals die earlier, survivors become a progressively more robust subset. Cohort mortality is therefore the mean hazard among those still alive: the individual Gompertz hazard multiplied by mean frailty among survivors,
\begin{equation}
    \bar\mu(x)=a\,e^{\,bx}\;\mathbb{E}\!\left[Z \mid T>x\right],
    \quad
    \mathbb{E}\!\left[Z \mid T>x\right]=\frac{1}{1+\gamma\,\dfrac{a}{b}\big(e^{\,bx}-1\big)}.
    \label{eq:mubar-decomp}
\end{equation}
The Gompertz factor $a\,e^{bx}$ rises without bound, whereas mean frailty among survivors falls as selection removes frailer individuals. Averaging over survivors gives the gamma--Gompertz hazard \cite{vaupel1979impact},
\begin{equation}
    \bar\mu(x \mid \theta)=
     \frac{a\,e^{\,bx}}
     {1+\gamma\,\dfrac{a}{b}\big(e^{\,bx}-1\big)},
    \quad \theta=(a,b,\gamma).
     \label{eq:ggm}
\end{equation}

Gompertz growth and selection jointly determine the cohort aging rate. Because $\mathrm{d}\log\bar\mu/\mathrm{d}x = b-\gamma\,\bar\mu(x)$, population mortality rises at the individual slope $b$ minus a selection term that grows with the hazard. As selection strengthens, this term approaches $b$ and the hazard approaches
\begin{equation}
    \pi=\lim_{x\to\infty}\bar\mu(x \mid \theta)=\frac{b}{\gamma}.
\label{eq:plateau}
\end{equation}
The plateau is not an individual ceiling; it is the population level at which selection offsets rising individual mortality. The hazard approaches $\pi$ but cannot decline. This monotonicity is a substantive restriction, chosen because age~65 lies within the senescent mortality regime. 

Across adult and old age, mortality rises approximately exponentially within cohorts. Historical declines across periods or cohorts instead lower mortality at a given age across calendar time or successive generations; they do not imply that mortality falls as a cohort ages \cite{zheng2016age,hansen2025decomposition}. This age pattern also has a mechanistic basis: models of biological aging derive Gompertz-like growth from the cumulative, self-reinforcing failure of interdependent components \cite{flietner2025unifying,nielsen2024gompertz}.

At the oldest ages, empirical studies support either continued Gompertz growth \cite{gavrilov2011mortality,gavrilova2015biodemography,gavrilov2019new,Danetal23} or a slowing rise followed by leveling \cite{thatcher1998force,gampe2010human,barbi2018plateau,belzile2021human}; they do not provide clear evidence of a sustained decrease. Our modeling framework represents both supported patterns: $\gamma=0$ yields continued Gompertz growth, whereas $\gamma>0$ allows selection to slow the cohort hazard toward a plateau. It does not include a declining phase. 

This restriction does not make a decline mathematically impossible: under other forms of heterogeneity, selection could make the aggregate cohort hazard fall even while every individual's hazard rises. Such a pattern lies outside the model and would be represented as a plateau.

\subsection*{Landmarks of the mortality schedule}\label{sec:onset}
From each fitted hazard, we derive four features of cohort mortality. The modal age at death marks the peak of the death distribution,
\begin{equation}
    x^{\mathrm{mode}}=\arg\max_{x}\,\bar\mu(x)\,\bar S(x),
    \qquad
    \bar S(x)=\exp\!\Big\{-\!\int_{0}^{x}\!\bar\mu(u)\,\mathrm{d}u\Big\}.
    \label{eq:mode}
\end{equation}
The other three describe late-life deceleration, which we locate with the life-table aging rate,
\begin{equation}
    k(x)=\frac{\mathrm{d}}{\mathrm{d}x}\log\bar\mu(x\mid\theta)=b-\gamma\,\bar\mu(x\mid\theta),
    \label{eq:lar}
\end{equation}
a standard measure of old-age deceleration \cite{horiuchi1990age,horiuchi1998deceleration}. Under a pure Gompertz law, $k=b$ at every age. Under the gamma--Gompertz model, $k$ declines from $b$ toward zero as selection strengthens. We define landmarks from turning points in this decline rather than from an external age threshold, allowing them to move with each cohort's schedule.

\paragraph{Age of mortality deceleration.}
We define $x^{\mathrm{dec}}$ as the age at which the decline in the aging rate accelerates most strongly, the minimum of $k''$,
\begin{equation}
x^{\mathrm{dec}}=\arg\min_{x}\, k''(x).
\label{eq:dec}
\end{equation}
At this point, mortality first departs sharply from Gompertz growth. For this model it falls at a fixed level of the aging rate, $k\approx0.79\,b$.

\paragraph{Onset of the plateau.}
After $x^{\mathrm{dec}}$, the aging rate continues to fall. We define plateau onset $x^{\mathrm{pl}}$ where this decline is fastest, the minimum of $k'$,
\begin{equation}
    x^{\mathrm{pl}}=\arg\min_{x} k'(x).
\label{eq:pl}
\end{equation}
By this age, the aging rate has halved to $k=b/2$. With $b\approx0.1$ across the fitted series, mortality at $x^{\mathrm{pl}}$ rises by about 5\% per year, compared with roughly 10\% in the exponential phase. This landmark marks entry into the near-flat tail identified as the plateau in recent studies \cite{barbi2018plateau,belzile2021human,belzile2022there}; it does not require an exactly constant hazard.

\paragraph{Level of the plateau.}
We summarize the height of the tail by the constant hazard equivalent to mortality beyond $x^{\mathrm{pl}}$. If $e(x^{\mathrm{pl}})$ denotes remaining life expectancy at plateau onset, the equivalent hazard is $1/e(x^{\mathrm{pl}})$ \cite{cohen2010life}.

\paragraph{Spacing of the landmarks.}
Because both landmarks sit at fixed levels of $k$, the gap between them depends only on the Gompertz slope\footnote{The minima fall at fixed values of $g(x)=1+\tfrac{\gamma a}{b-\gamma a}e^{bx}$---namely $g=3-\sqrt3$ (where $k=0.79\,b$) and $g=2$ (where $k=b/2$)---so subtracting the two ages cancels $a$ and $\gamma$.},
\begin{equation}
    x^{\mathrm{pl}}-x^{\mathrm{dec}} = \frac{1}{b}\,\ln\!\frac{1}{2-\sqrt{3}}
     \approx \frac{1.3169}{b}.
    \label{eq:spacing}
\end{equation}

The interval widens as $b$ falls and narrows as $b$ rises, regardless of $a$ or $\gamma$. Because $b$ changes little between adjacent cohorts \cite{patricio2026rhythm}, the interval usually remains stable as the schedule shifts later.

Within this model, the equivalence is exact. Any widening, narrowing, or sex divergence in the deceleration-to-plateau interval corresponds to a change in the fitted slope~$b$ (Eq.~\eqref{eq:spacing}). Selection, historical shocks, and sex differences are possible mechanisms that may distort $b$, not separate determinants of the interval.

Two mechanisms may shape the fitted slope. Selection before age~65 determines who enters the observation window; heavier prior selection can produce a flatter fitted slope even if individual aging is unchanged \cite{steinsaltz2006understanding,alter1989frailty}. Wars and epidemics can also affect several cohorts at once and shift their fitted rates of aging \cite{patricio2026rhythm}. By Eq.~\eqref{eq:spacing}, any associated widening of the interval appears as a lower $b$.

The same logic applies to the sex contrast. Higher male mortality at younger and middle ages \cite{beltransanchez2015excess} may leave more selected survivors and a flatter fitted slope. The wider male interval is therefore consistent with either a lower late-life slope or stronger earlier selection; these data cannot distinguish them. Comparative evidence across mammals likewise suggests that adversity often changes mortality levels more than senescence rates \cite{colchero2021long,lemaitre2020sex}.

\subsection*{Likelihood}
At the oldest ages, finite cohorts, few deaths, residual heterogeneity, and registration irregularities can make death counts more variable than a Poisson model allows \cite{wong2026bayesian}. We therefore use a negative-binomial likelihood with overdispersion parameter $\phi$; it approaches the Poisson model as $\phi\to\infty$. With $\theta=(a,b,\gamma)$ the cohort hazard parameters:
\begin{align}
    D_{x,j}\mid \theta_j,\phi
    &\sim \operatorname{NegBinomial}\!\big(E_{x,j}\,\bar\mu(x \mid \theta_j),\ \phi\big),
    \label{eq:nb}\\[2pt]
    \mathbb{VAR}\!\left[D_{x,j}\right]
     &= E_{x,j}\,\bar\mu(x \mid \theta_j)
     + \frac{\big(E_{x,j}\,\bar\mu(x \mid \theta_j)\big)^{2}}{\phi},
    \label{eq:nb-var}
\end{align}
where $D_{x,j}$ is the death count at age $x$ in cohort $j$, $E_{x,j}$ is the exposure to death, and the mean is $\mathbb{E}[D_{x,j}]=E_{x,j}\,\bar\mu(x\mid\theta_j)$.

\subsection*{Cohort dynamics and priors}
Adjacent birth cohorts experience nearly the same historical  periods, and the shocks that shape old-age mortality---wars, epidemics, gains in nutrition and medicine---act on them alike and accumulate rather than cancel. Their mortality schedules are therefore strongly related. Fitting each cohort independently would discard this structure and rely heavily on sparse deaths at the oldest ages. Cohort-specific estimates show nonstationary levels but approximately stationary first differences, consistent with a random walk \cite{patricio2026rhythm}.

We let each parameter follow a first-order random walk across cohorts. This structure links neighbors and stabilizes estimates at sparse ages without imposing a fixed trend. The log scale keeps all parameters positive and represents proportional cohort changes as additive steps,
\begin{align}
    \log a_{j} &\sim \operatorname{Normal}(\log a_{j-1},\ \tau_a),\\
    \log b_{j} &\sim \operatorname{Normal}(\log b_{j-1},\ \tau_b),\\
    \log \gamma_{j} &\sim \operatorname{Normal}(\log \gamma_{j-1},\ \tau_{\gamma}),
    \label{eq:rw}
\end{align}
for $j=2,\dots,J$, with the first cohort given a flat prior and identified by its own data.

Each walk scale measures the typical parameter change between adjacent cohorts. We assign weakly informative half-normal priors to the three scales and to overdispersion,
\begin{equation}
    \tau_a,\tau_b,\tau_{\gamma}\sim \operatorname{Half\text{-}Normal}(0,0.5),
    \qquad
    \phi \sim \operatorname{Half\text{-}Normal}(0,1),
    \label{eq:hyper}
\end{equation}
which favor small steps while allowing the data to determine their size and keep overdispersion positive while retaining the Poisson limit \cite{gelman2006prior}. We also restrict $a\le1$ and $\gamma\le1$. These bounds contain plausible values for the study populations and improve the sampler's convergence without constraining the posterior estimates.

\begin{figure}[p]
    \centering
    \includegraphics[width=\textwidth]{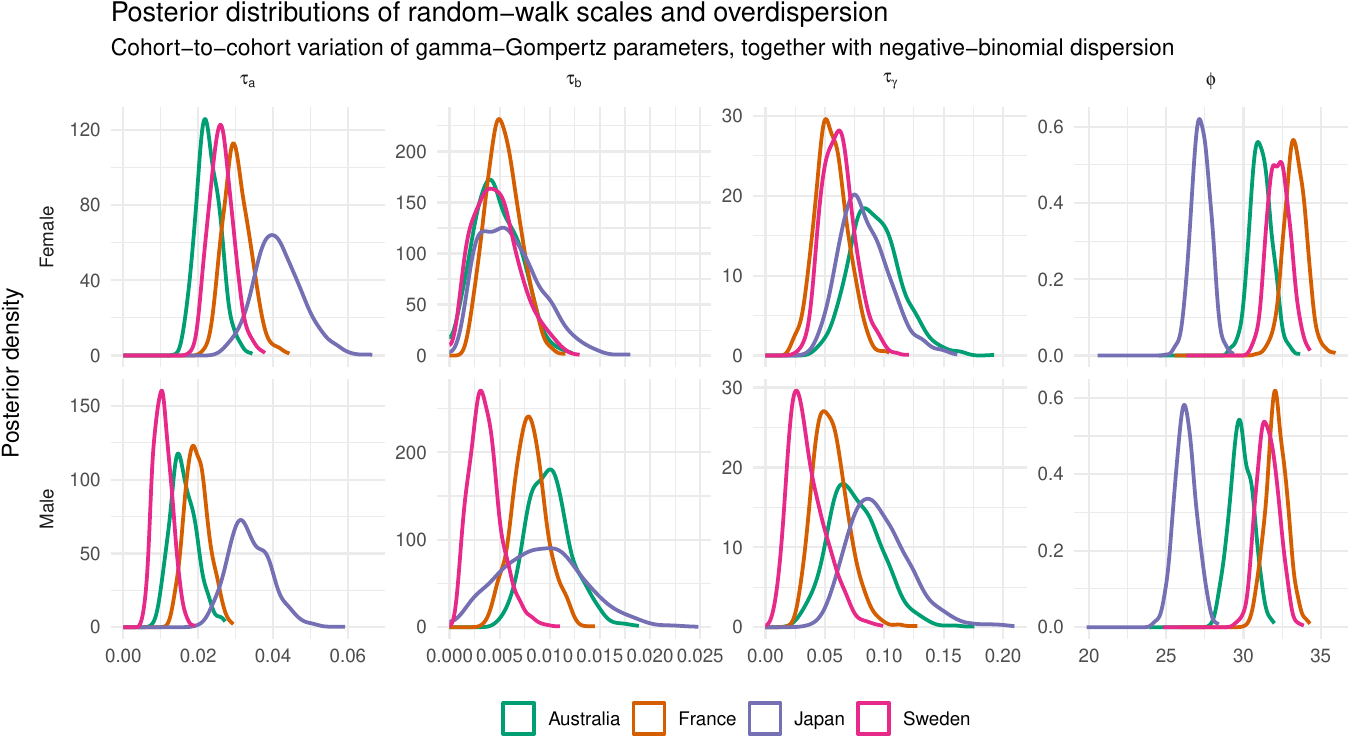}
    \caption{\textbf{Posterior distributions of random-walk scales and negative-binomial dispersion.} Kernel-density estimates show the posterior distributions of the random-walk scales $\tau_a$, $\tau_b$, and $\tau_\gamma$ and the negative-binomial dispersion parameter $\phi$ for Australia, France, Japan, and Sweden, separately for females (top) and males (bottom). The random-walk scales measure the typical cohort-to-cohort change in $\log a$, $\log b$, and $\log\gamma$; smaller values correspond to smoother cohort trajectories. Larger values of $\phi$ indicate less residual variation in death counts, with the Poisson limit approached as $\phi\rightarrow\infty$. Colors identify populations. Curves summarize 8{,}000 retained posterior draws from each fit.}
    \label{fig:posterior_hyperparameters}
\end{figure}

The posterior distributions provide a direct check on how strongly the data update these hyperparameters. In four illustrative populations, all three random-walk scales are sharply concentrated relative to their $\operatorname{Half\text{-}Normal}(0,0.5)$ prior scale (Fig.~\ref{fig:posterior_hyperparameters}). The posterior distributions of $\phi$ lie between approximately 25 and 34, far above the scale favored by its $\operatorname{Half\text{-}Normal}(0,1)$ prior. The likelihood therefore provides substantial information about the cohort-to-cohort variation and residual dispersion in each population--sex series.

\subsection*{Posterior sampling}
The hazard \eqref{eq:ggm}, likelihood \eqref{eq:nb}, and cohort priors \eqref{eq:rw}--\eqref{eq:hyper} form one joint model. Joint estimation propagates parameter uncertainty to every derived landmark. We use the No-U-Turn sampler in Stan \cite{stan} with four chains, each retaining 2{,}000 iterations after 4{,}000 warm-up iterations, for 8{,}000 posterior draws. All parameters had $\widehat{R}<1.05$, and no chain produced divergent transitions. 

\subsection*{Model checking}
\begin{figure}[p]
  \centering
  \includegraphics[width=\textwidth]{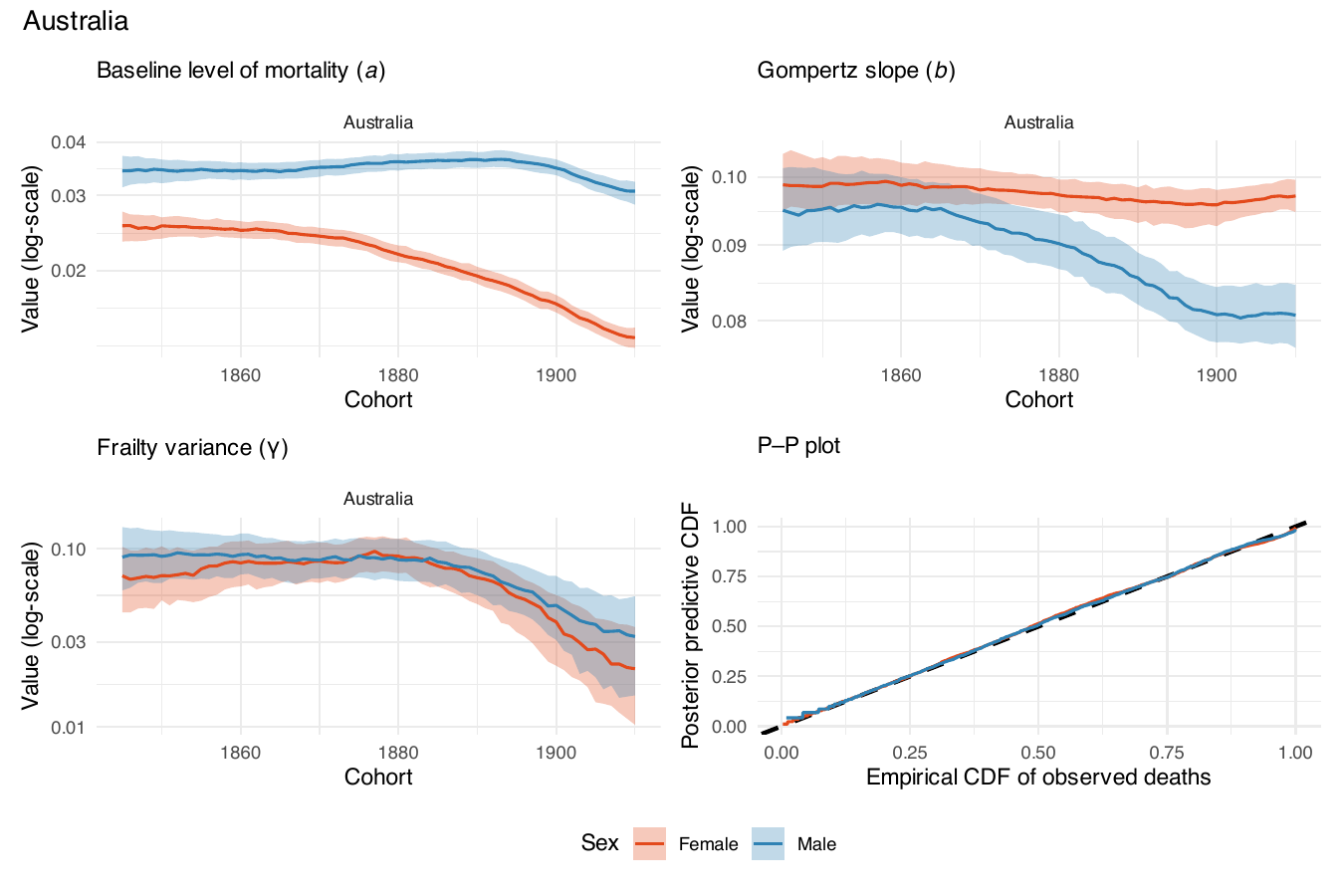}
    \caption{\textbf{Model diagnostics for Australia.} The first three panels show posterior trajectories of baseline mortality ($a$), the Gompertz slope ($b$), and frailty variance ($\gamma$) on logarithmic scales. Lines give posterior mode and bands 95\% credible intervals. The fourth panel compares posterior predictive and observed death-count distributions; agreement follows the dashed diagonal. Female estimates are red and male estimates blue.}
    \label{fig:fit_australia}
\end{figure}

\begin{figure}[p]
\centering
\includegraphics[width=\textwidth]{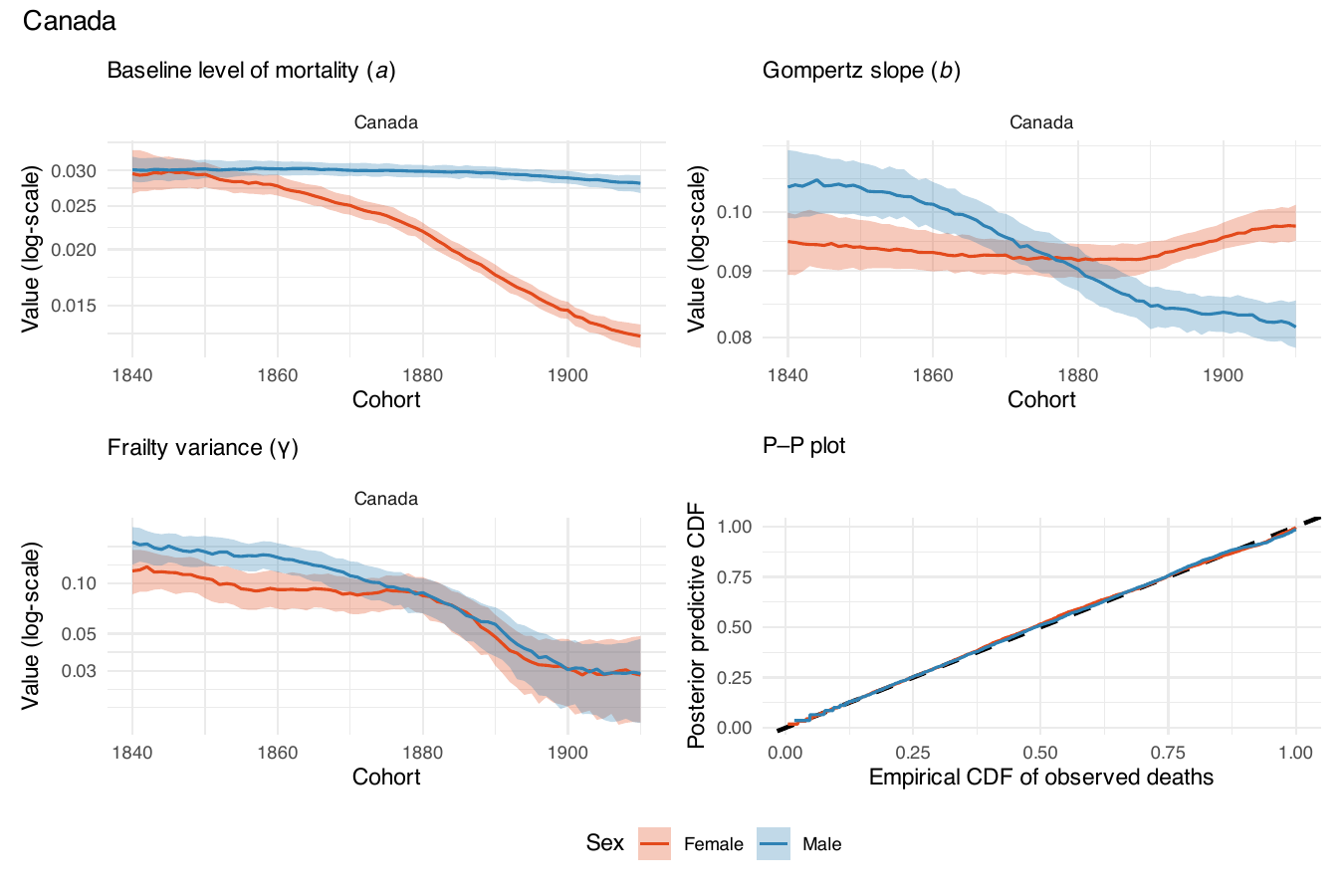}
\caption{\textbf{Model diagnostics for Canada.} The first three panels show posterior trajectories of baseline mortality ($a$), the Gompertz slope ($b$), and frailty variance ($\gamma$) on logarithmic scales. Lines give posterior mode and bands 95\% credible intervals. The fourth panel compares posterior predictive and observed death-count distributions; agreement follows the dashed diagonal. Female estimates are red and male estimates blue.}
\label{fig:fit_Canada}
\end{figure}

\begin{figure}[p]
\centering
\includegraphics[width=\textwidth]{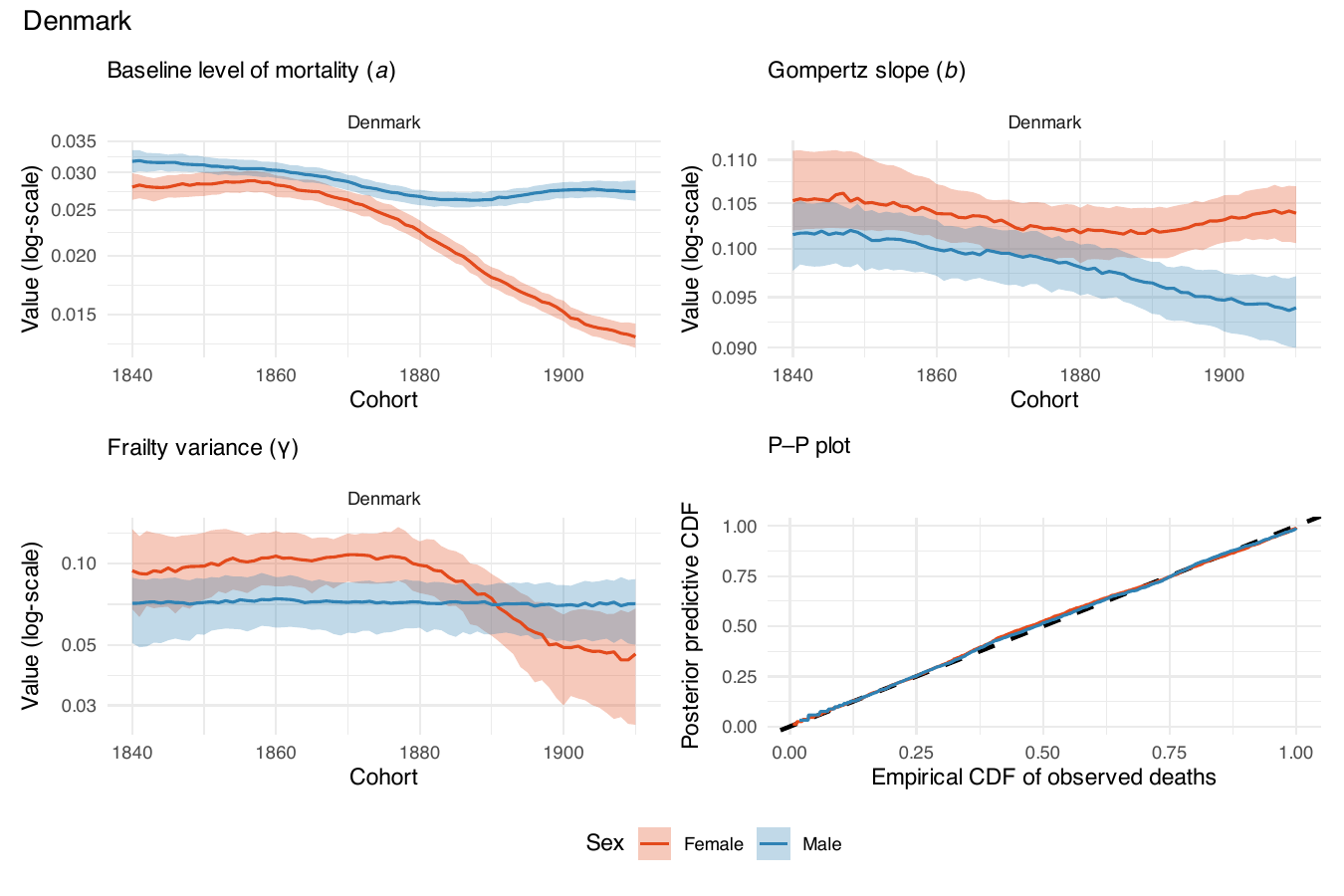}
\caption{\textbf{Model diagnostics for Denmark.}  The first three panels show posterior trajectories of baseline mortality ($a$), the Gompertz slope ($b$), and frailty variance ($\gamma$) on logarithmic scales. Lines give posterior mode and bands 95\% credible intervals. The fourth panel compares posterior predictive and observed death-count distributions; agreement follows the dashed diagonal. Female estimates are red and male estimates blue.}
\label{fig:fit_Denmark}
\end{figure}

\begin{figure}[p]
\centering
\includegraphics[width=\textwidth]{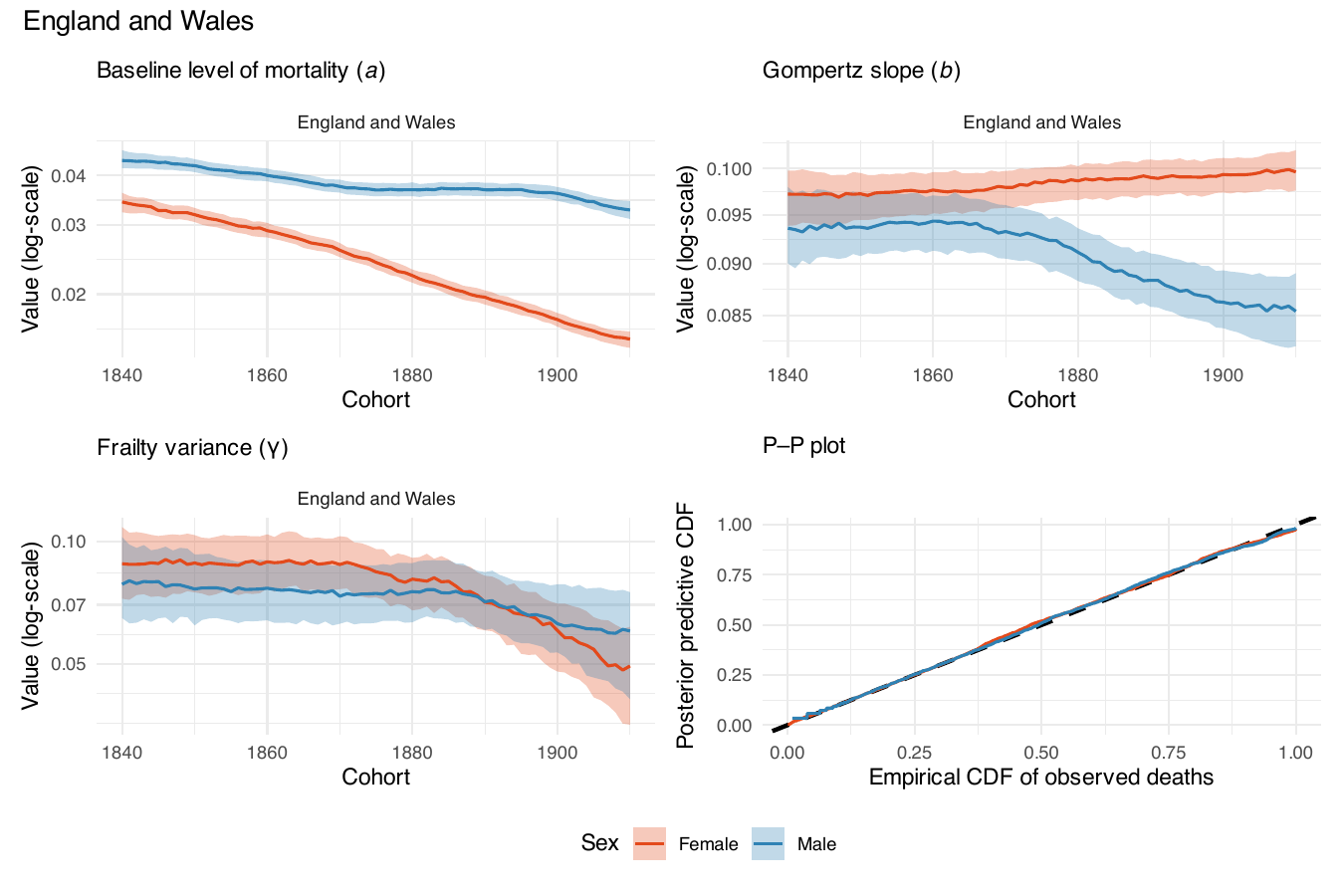}
\caption{\textbf{Model diagnostics for England and Wales.} The first three panels show posterior trajectories of baseline mortality ($a$), the Gompertz slope ($b$), and frailty variance ($\gamma$) on logarithmic scales. Lines give posterior mode and bands 95\% credible intervals. The fourth panel compares posterior predictive and observed death-count distributions; agreement follows the dashed diagonal. Female estimates are red and male estimates blue.}
\label{fig:fit_England and Wales}
\end{figure}

\begin{figure}[p]
\centering
\includegraphics[width=\textwidth]{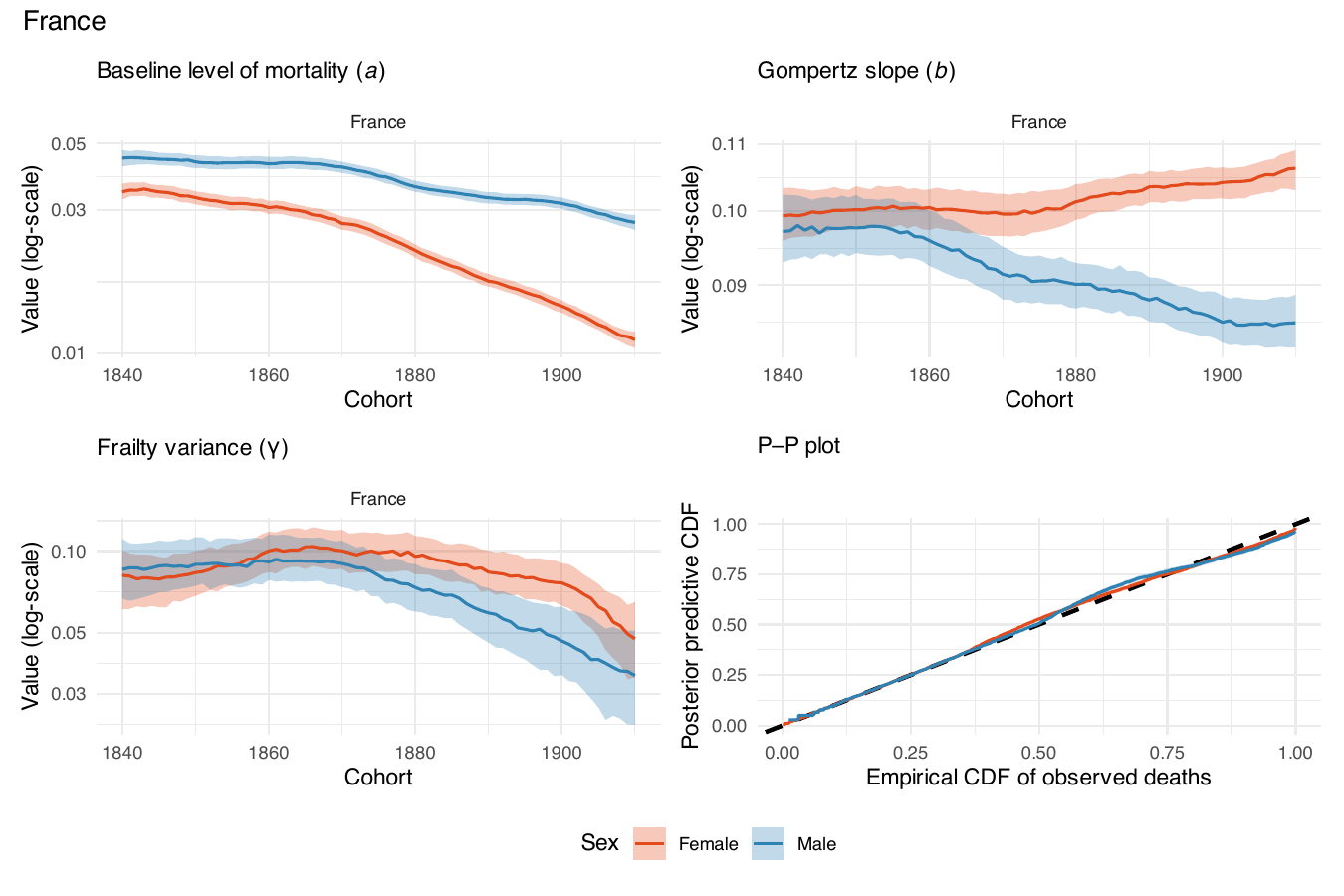}
\caption{\textbf{Model diagnostics for France.} The first three panels show posterior trajectories of baseline mortality ($a$), the Gompertz slope ($b$), and frailty variance ($\gamma$) on logarithmic scales. Lines give posterior mode and bands 95\% credible intervals. The fourth panel compares posterior predictive and observed death-count distributions; agreement follows the dashed diagonal. Female estimates are red and male estimates blue.}
\label{fig:fit_France}
\end{figure}

\begin{figure}[p]
\centering
\includegraphics[width=\textwidth]{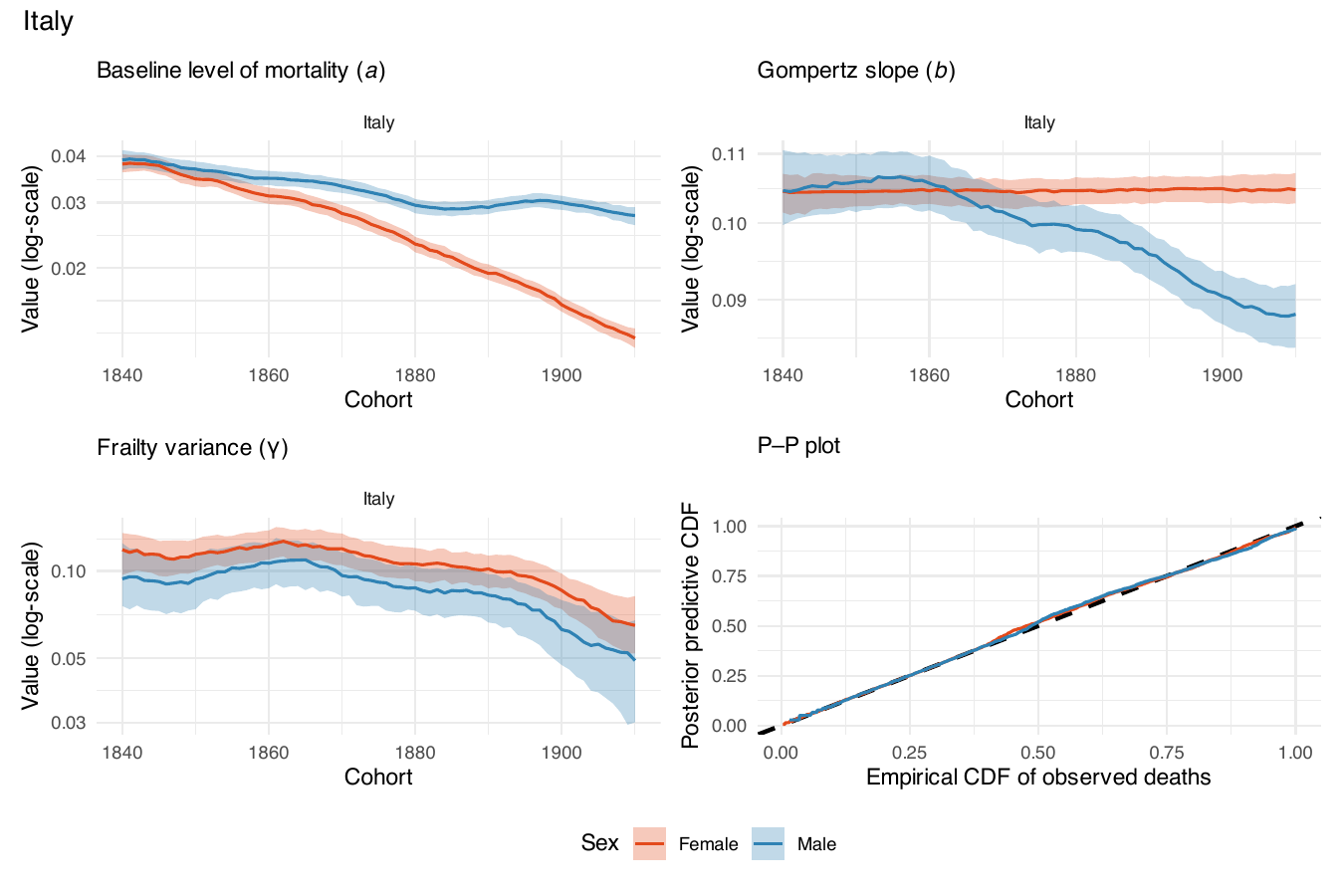}
\caption{\textbf{Model diagnostics for Italy.} The first three panels show posterior trajectories of baseline mortality ($a$), the Gompertz slope ($b$), and frailty variance ($\gamma$) on logarithmic scales. Lines give posterior mode and bands 95\% credible intervals. The fourth panel compares posterior predictive and observed death-count distributions; agreement follows the dashed diagonal. Female estimates are red and male estimates blue.}
\label{fig:fit_Italy}
\end{figure}

\begin{figure}[p]
\centering
\includegraphics[width=\textwidth]{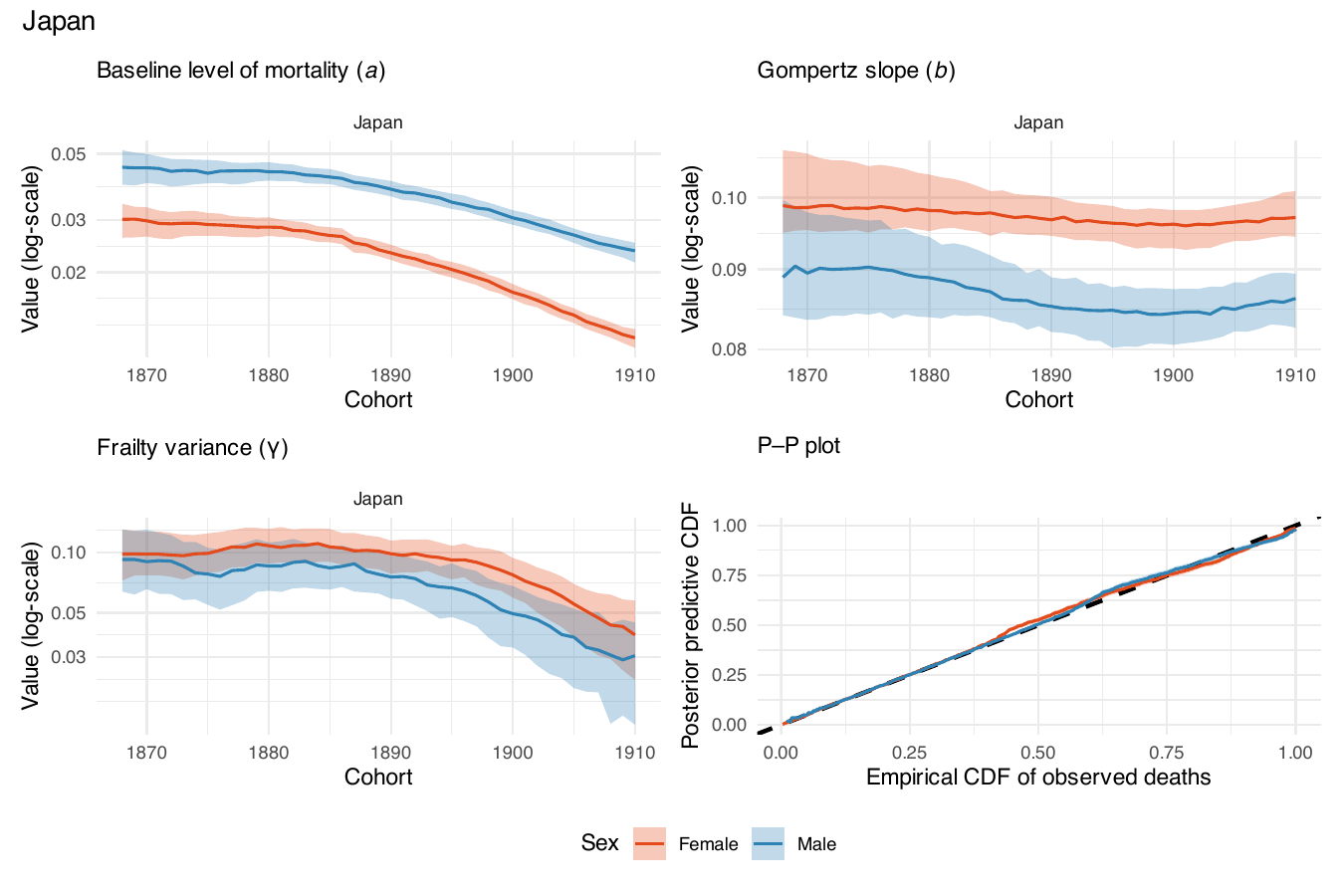}
\caption{\textbf{Model diagnostics for Japan.} The first three panels show posterior trajectories of baseline mortality ($a$), the Gompertz slope ($b$), and frailty variance ($\gamma$) on logarithmic scales. Lines give posterior mode and bands 95\% credible intervals. The fourth panel compares posterior predictive and observed death-count distributions; agreement follows the dashed diagonal. Female estimates are red and male estimates blue.}
\label{fig:fit_Japan}
\end{figure}

\begin{figure}[p]
\centering
\includegraphics[width=\textwidth]{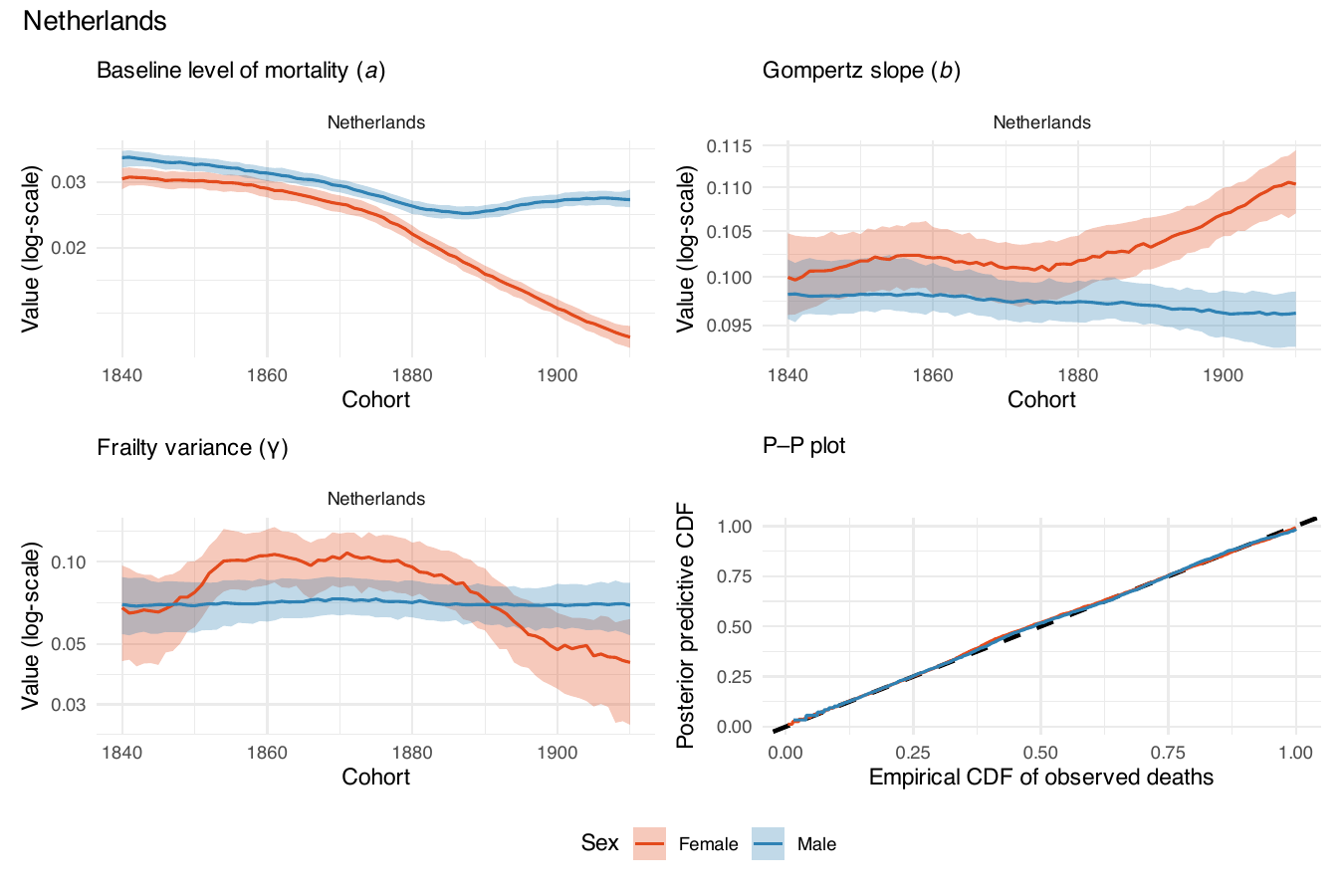}
\caption{\textbf{Model diagnostics for Netherlands.} The first three panels show posterior trajectories of baseline mortality ($a$), the Gompertz slope ($b$), and frailty variance ($\gamma$) on logarithmic scales. Lines give posterior mode and bands 95\% credible intervals. The fourth panel compares posterior predictive and observed death-count distributions; agreement follows the dashed diagonal. Female estimates are red and male estimates blue.}
\label{fig:fit_Netherlands}
\end{figure}

\begin{figure}[p]
\centering
\includegraphics[width=\textwidth]{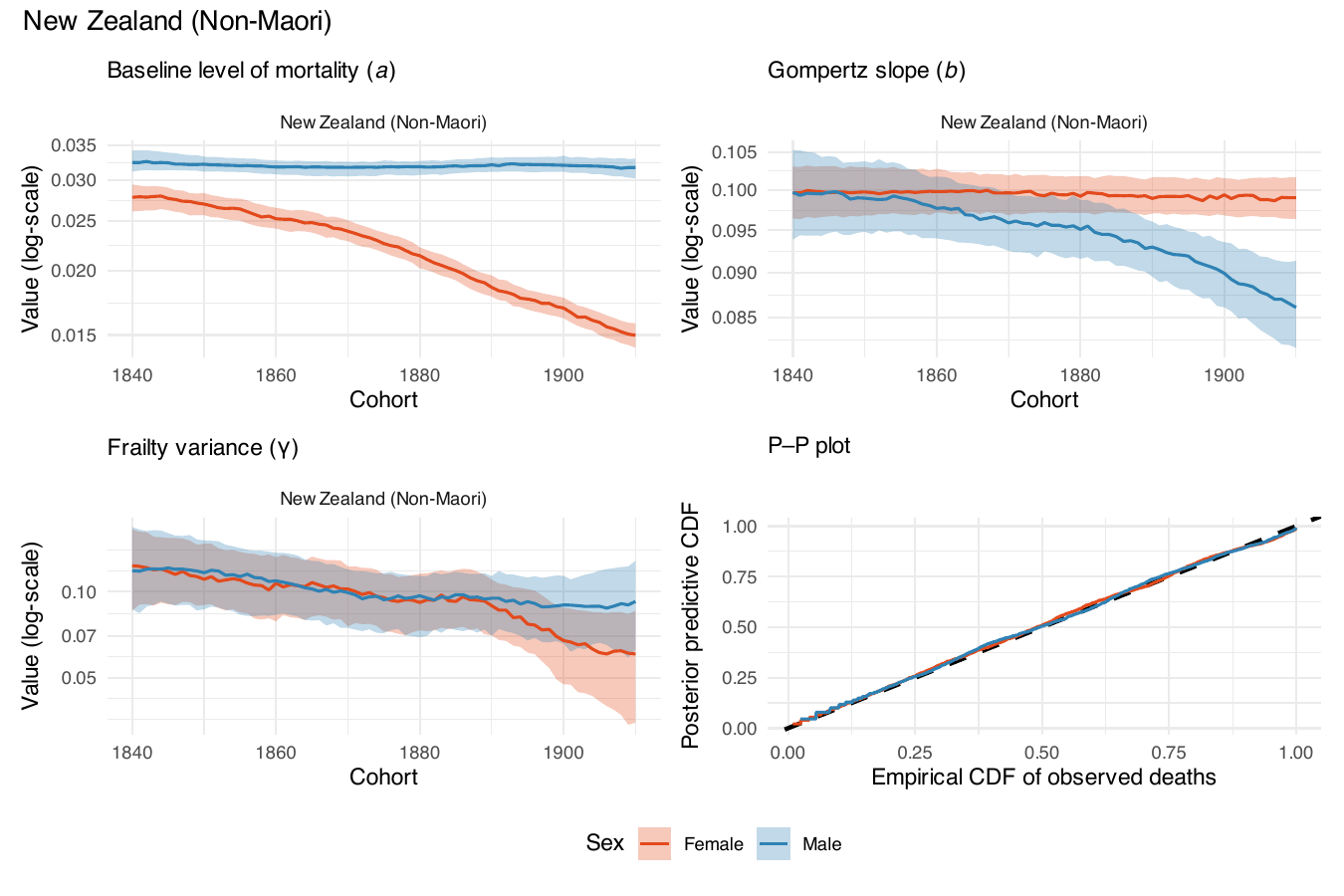}
\caption{\textbf{Model diagnostics for New Zealand (Non-Maori).} The first three panels show posterior trajectories of baseline mortality ($a$), the Gompertz slope ($b$), and frailty variance ($\gamma$) on logarithmic scales. Lines give posterior mode and bands 95\% credible intervals. The fourth panel compares posterior predictive and observed death-count distributions; agreement follows the dashed diagonal. Female estimates are red and male estimates blue.}
\label{fig:fit_New_Zealand}
\end{figure}

\begin{figure}[p]
\centering
\includegraphics[width=\textwidth]{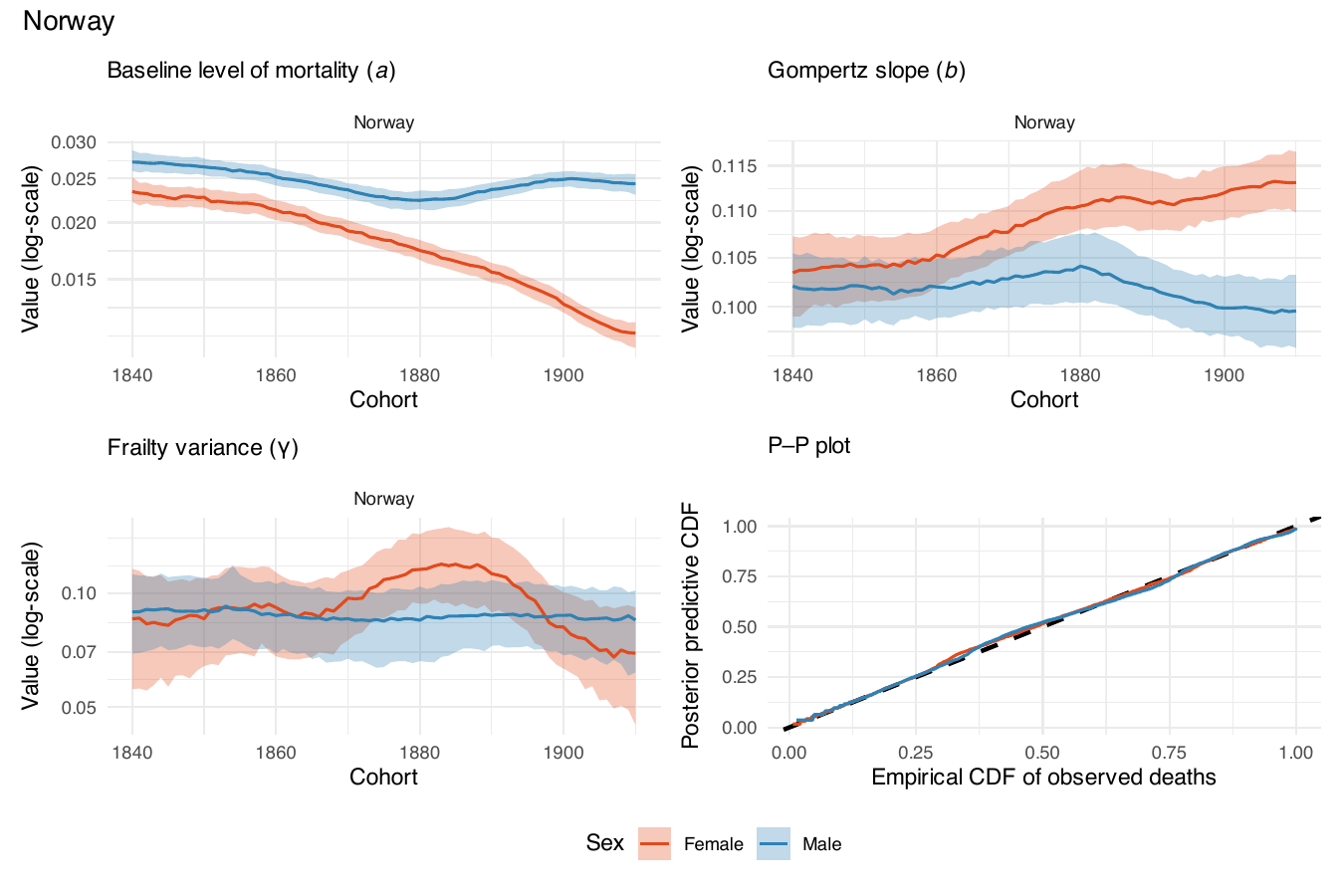}
\caption{\textbf{Model diagnostics for Norway.}The first three panels show posterior trajectories of baseline mortality ($a$), the Gompertz slope ($b$), and frailty variance ($\gamma$) on logarithmic scales. Lines give posterior mode and bands 95\% credible intervals. The fourth panel compares posterior predictive and observed death-count distributions; agreement follows the dashed diagonal. Female estimates are red and male estimates blue.}
\label{fig:fit_Norway}
\end{figure}

\begin{figure}[p]
\centering
\includegraphics[width=\textwidth]{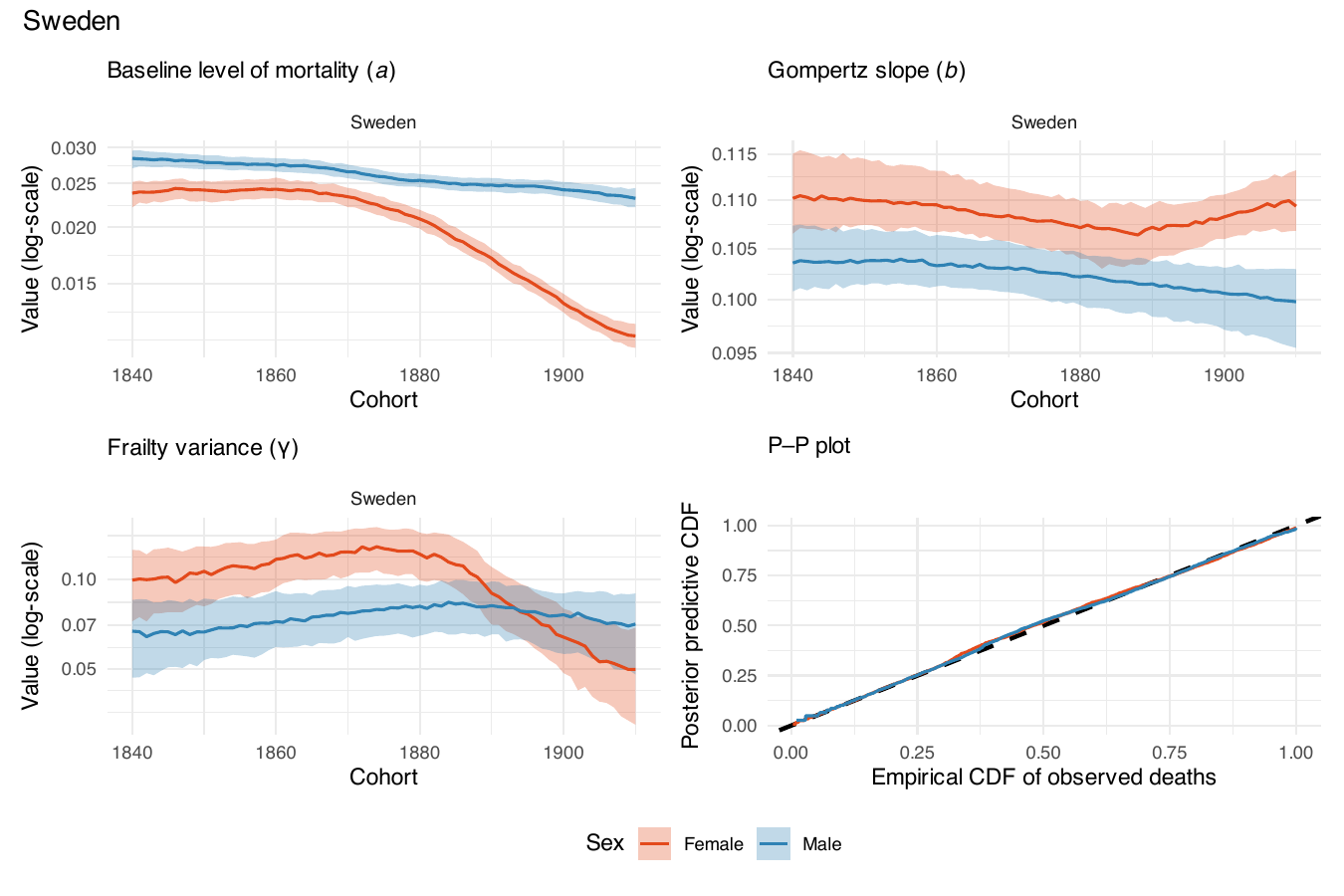}
\caption{\textbf{Model diagnostics for Sweden.} The first three panels show posterior trajectories of baseline mortality ($a$), the Gompertz slope ($b$), and frailty variance ($\gamma$) on logarithmic scales. Lines give posterior mode and bands 95\% credible intervals. The fourth panel compares posterior predictive and observed death-count distributions; agreement follows the dashed diagonal. Female estimates are red and male estimates blue.}
\label{fig:fit_Sweden}
\end{figure}

\begin{figure}[p]
\centering
\includegraphics[width=\textwidth]{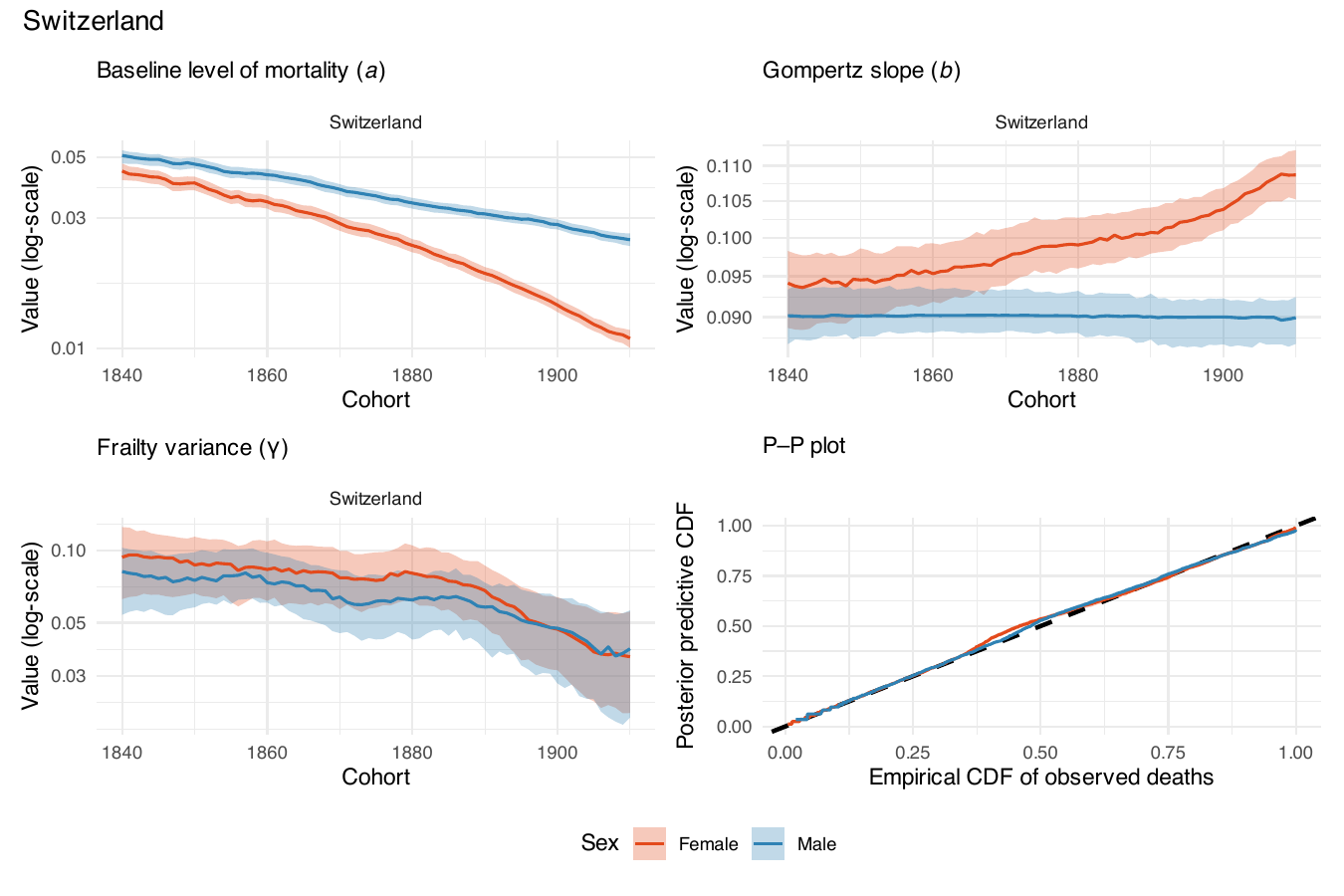}
\caption{\textbf{Model diagnostics for Switzerland.} The first three panels show posterior trajectories of baseline mortality ($a$), the Gompertz slope ($b$), and frailty variance ($\gamma$) on logarithmic scales. Lines give posterior mode and bands 95\% credible intervals. The fourth panel compares posterior predictive and observed death-count distributions; agreement follows the dashed diagonal. Female estimates are red and male estimates blue.}
\label{fig:fit_Switzerland}
\end{figure}

We assess fit with posterior predictive checks. For each draw, we simulate death counts and compare them with observations across ages and cohorts. The probability--probability plots follow the identity line closely in every population (Figs.~\ref{fig:fit_australia}--\ref{fig:fit_Switzerland}); the same figures show the cohort trajectories of $a$, $b$, and $\gamma$.

\subsection*{Testing for the plateau}
The gamma--Gompertz hazard has a finite limit, $\pi=b/\gamma$, whenever $\gamma>0$ (Eq.~\eqref{eq:plateau}). As $\gamma\downarrow 0$, the selection term in Eq.~\eqref{eq:lar} vanishes, the aging rate $k(x)=b$, and the hazard reduces to pure Gompertz model \cite{gompertz1825xxiv,vaupel1979impact}, with neither deceleration nor a plateau. We can therefore test for a plateau within the same model by asking whether the data support $\gamma>0$.

We implement this test with a spike-and-slab prior on frailty variance \cite{mitchell1988bayesian,george1993variable}. For each cohort $j$, the spike fixes $\gamma_j=0$ and produces a pure Gompertz tail; the slab keeps $\gamma_j$ positive and follows the same random walk as the continuous model:
\begin{equation}
    \gamma_j =
    \begin{cases}
        0, & \text{with prior probability } w,\\[2pt]
        \gamma_j^{\mathrm{slab}}, & \text{with prior probability } 1-w,
    \end{cases}
    \qquad
    \log\gamma_j^{\mathrm{slab}} \sim \operatorname{Normal}\!\big(\log\gamma_{j-1}^{\mathrm{slab}},\ \tau_\gamma\big).
    \label{eq:spikeslab}
\end{equation}

The random walk within the slab preserves information sharing across cohorts, while the mixture compares the plateau and no-plateau states. The weight $w$ is shared across cohorts and given a uniform prior, $w\sim\operatorname{Beta}(1,1)$, which gives the two states equal prior weight on average. All other parameters, the likelihood~\eqref{eq:nb}, and the sampler remain unchanged.

Each cohort enters the likelihood through its own $\theta_j$, so we evaluate the two states separately and integrate out the discrete indicator, which Hamiltonian Monte Carlo cannot sample \cite{stan}. Define
\begin{equation}
    \mathcal{L}_j(\gamma)=\prod_x \operatorname{NegBinomial}\!\big(D_{x,j}\mid E_{x,j}\,\bar\mu(x\mid a_j,b_j,\gamma),\ \phi\big)
\end{equation}
as the cohort likelihood at a given frailty variance. 

The cohort contributes the mixture $w\,\mathcal{L} _j(0)+(1-w)\,\mathcal{L} _j(\gamma_j^{\mathrm{slab}})$ to the posterior. On each draw, its probability of belonging to the spike is
\begin{equation}
    \mathbb{P}\big(\gamma_j=0 \mid data\big)
    = \frac{w\,\mathcal{L} _j(0)}{\,w\,\mathcal{L} _j(0)+(1-w)\,\mathcal{L} _j\!\big(\gamma_j^{\mathrm{slab}}\big)\,},
    \label{eq:pnp}
\end{equation}
whose posterior mean is the marginal probability that cohort $j$ has no plateau. We compare this probability with the prior weight $w$. Values far below $w$ favor a plateau; a movement toward $w$ indicates weaker separation between the states.

\begin{figure}[htbp]
    \centering
    \includegraphics[width=\linewidth]{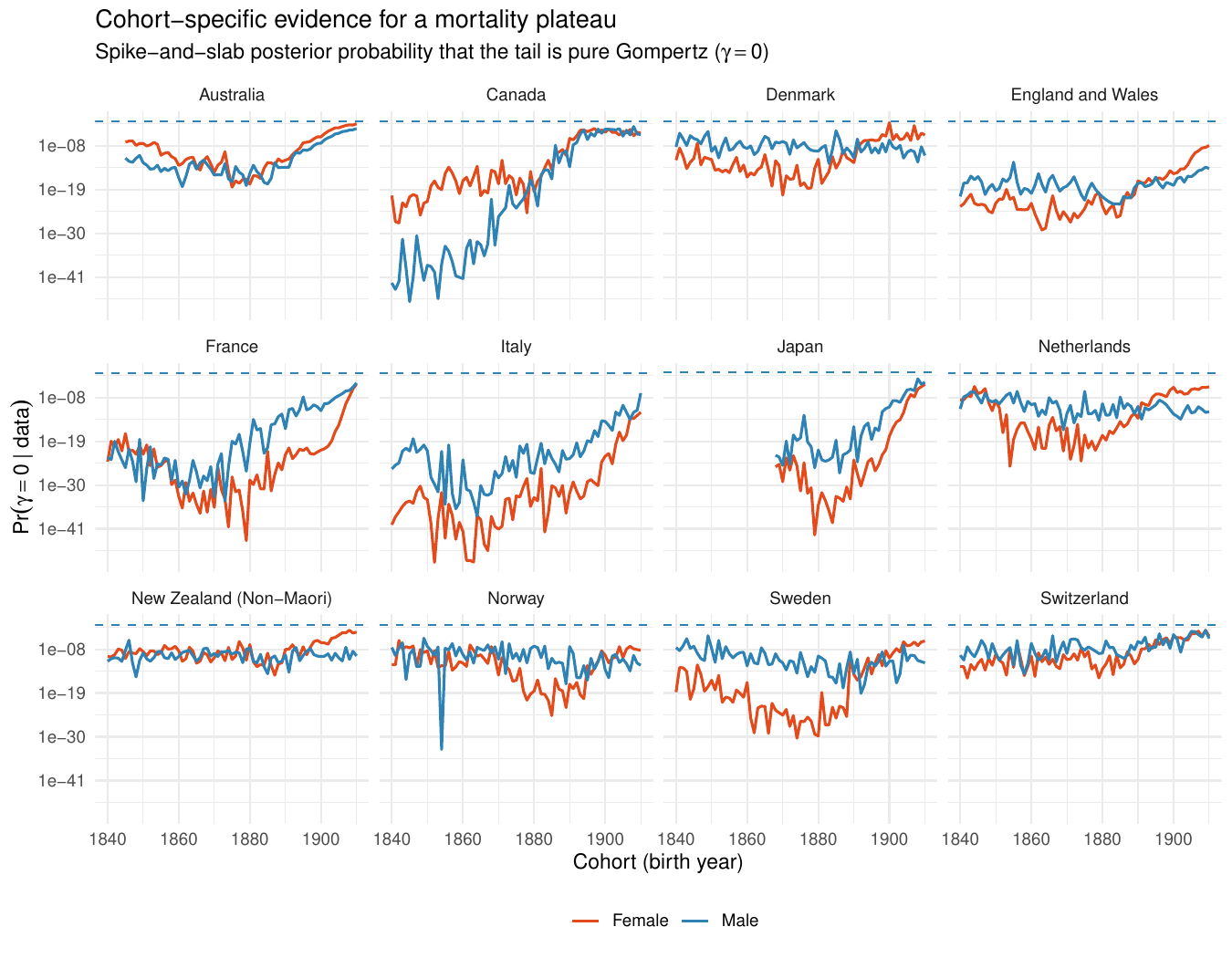}
        \caption{\textbf{Cohort-specific evidence for a mortality plateau.} Posterior probability that each cohort follows a pure Gompertz tail ($\gamma=0$, no plateau), estimated with the spike-and-slab model and shown on a logarithmic scale. Cohorts span 1840--1910, except in Australia (1845--1910) and Japan (1868--1910), across twelve populations. Female trajectories are red and male trajectories blue; dashed lines mark the prior weights. All trajectories remain below the prior benchmark, favoring a plateau. The rise in some recent cohorts indicates weaker, but still clear, support as follow-up becomes incomplete at the highest ages.}

    \label{fig:spike_slab}
\end{figure}

For every cohort, the posterior probability of no plateau lies far below $w$ (Fig.~\ref{fig:spike_slab}). It rises in some of the youngest cohorts but remains well below the prior benchmark. These cohorts have more deaths at many ages already reached, but their mortality histories remain incomplete at the highest ages, which weakens the contrast between plateau and no-plateau states. The spike-and-slab comparison therefore supports a plateau across the cohort range. We derive the landmarks from the continuous gamma--Gompertz fit, where $\gamma>0$ on every draw and each onset is defined.
 
\subsection*{Data}
We apply the model to Human Mortality Database cohort data \cite{hmd} from twelve low-mortality countries: the Nordic countries (Denmark, Norway, Sweden), the Netherlands, England and Wales, France, Italy, and Switzerland in Western and Southern Europe, and Australia, Canada, Japan, and New Zealand elsewhere. These populations provide long cohort series and continuous death registration, both needed to estimate moving late-life landmarks.

We fit each of the twenty-four population--sex series from age~65 onward, where senescent mortality dominates. Table~\ref{tab:cohorts} lists the cohorts in each series.

\subsection*{Code and data availability}
The data come from the publicly available Human Mortality Database
(\url{https://www.mortality.org}). The full analysis---data preparation, the Stan model, posterior sampling, landmark calculations, and all
figures---is openly available at
\begin{center}
    \url{https://github.com/scpatricio/dynamic_shifts_in_mortality_patterns}
\end{center}
so that results can be reproduced.

\clearpage 

\section*{Supplementary Figures}

\clearpage

\begin{figure}[p]
 \centering
 \includegraphics[width=.95\textwidth]{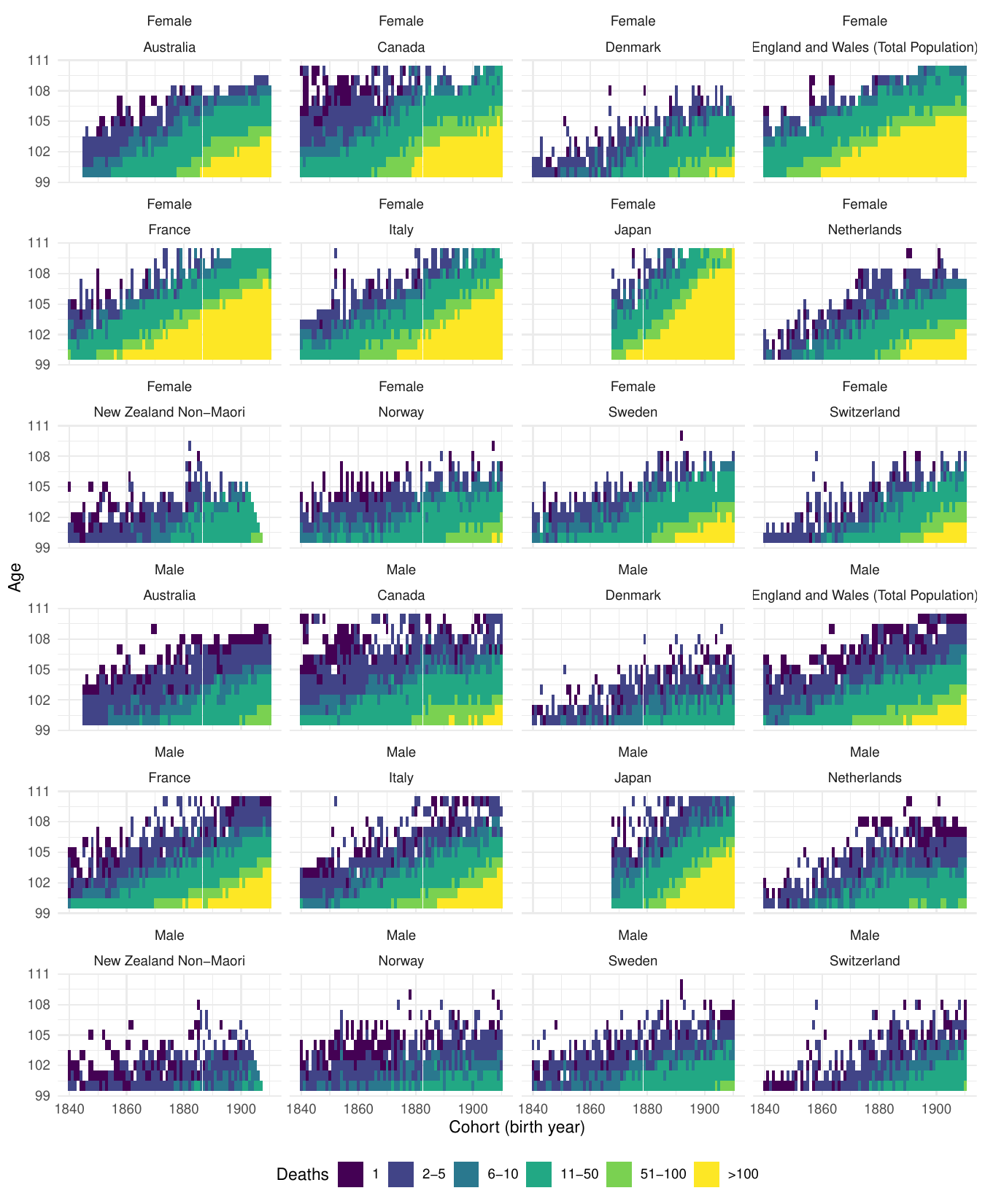}
  \caption{\textbf{Observed deaths at extreme ages, by age, cohort, and country.} Each cell gives the binned number of deaths at a single age above~100, by sex and country. Cohorts span 1840--1910, except in Australia (1845--1910) and Japan (1868--1910). The upper edge of the colored region marks the highest observed age in each cohort and moves upward as survival extends further into the tail. Above about age~108, most populations record fewer than ten deaths per age--cohort cell.}
 \label{fig:counts}
\end{figure}

\clearpage

\begin{figure}[p]
    \centering
    \includegraphics[width=\linewidth]{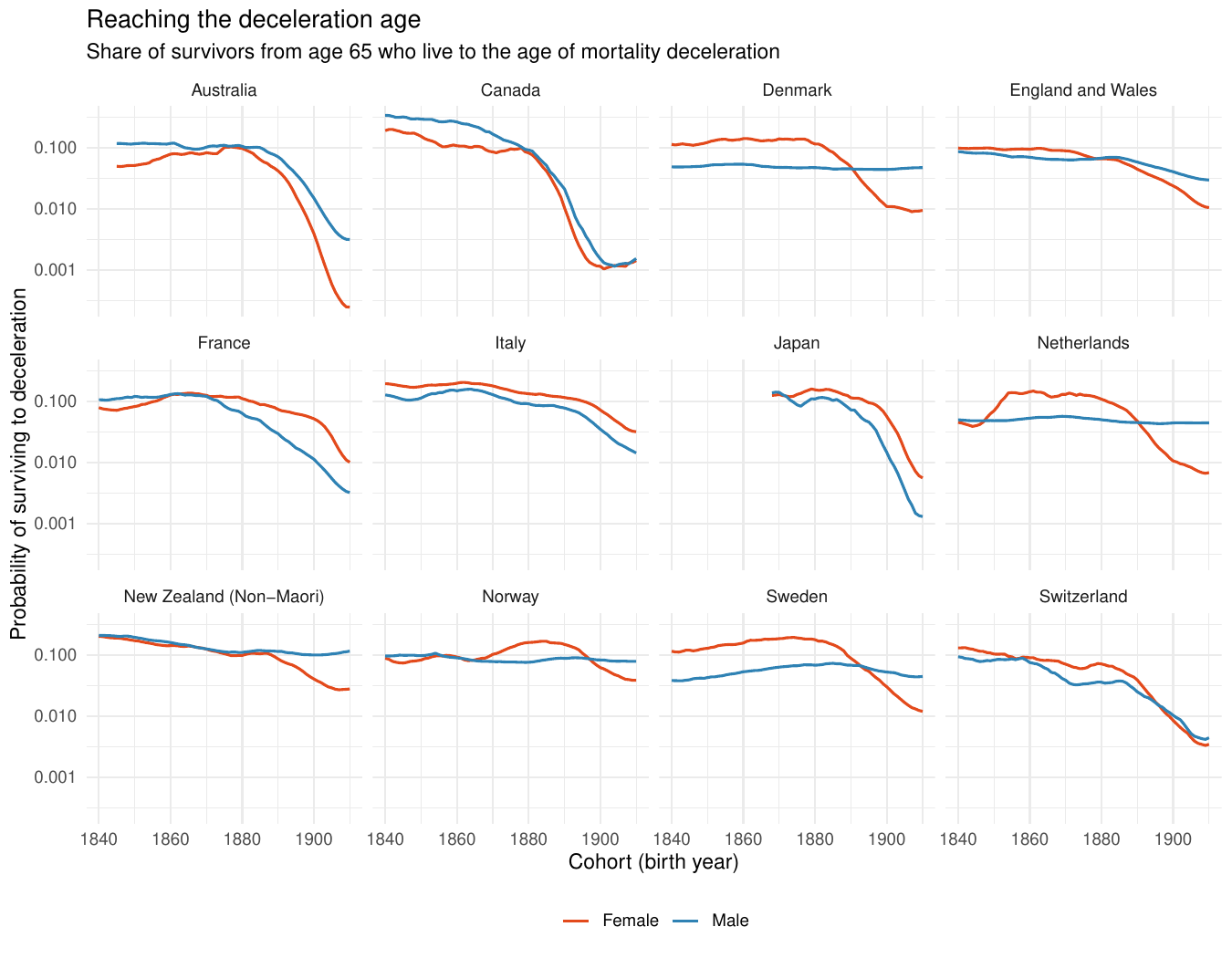}
    \caption{\textbf{Share of survivors reaching the age of mortality deceleration.} For each cohort, the figure shows on a logarithmic scale the posterior probability that an individual alive at age~65 survives to the cohort-specific age of mortality deceleration. Cohorts span 1840--1910, except in Australia (1845--1910) and Japan (1868--1910), across twelve populations.}
    \label{fig:sirvival}
\end{figure}

\clearpage

\clearpage

\clearpage

\clearpage

\clearpage

\clearpage

\clearpage

\clearpage

\clearpage

\clearpage

\clearpage

\clearpage

\clearpage

\clearpage 

\section*{Supplementary Tables}

\begin{table}[!htb]
 \centering
  \caption{\label{tab:mean_steps}Mean annual cohort change in the three mortality landmarks (years), by population and sex. Values give the average change between successive birth cohorts in the modal age at death, age of mortality deceleration, and plateau onset.}
  \centering
  \begin{tabular}[t]{llrrr}
    \hline
    Country & Sex & Mode & Deceleration & Plateau onset\\
    \hline
    Australia & Female & 0.0984 & 0.2850 & 0.2886\\
    & Male & 0.0234 & 0.2573 & 0.2953\\
    Canada & Female & 0.1268 & 0.3280 & 0.3220\\
    & Male & 0.0223 & 0.3524 & 0.4031\\
    Denmark & Female & 0.1036 & 0.2103 & 0.2143\\
    & Male & 0.0237 & 0.0405 & 0.0564\\
    England and Wales & Female & 0.1180 & 0.1986 & 0.1929\\
    & Male & 0.0466 & 0.1159 & 0.1351\\
    France & Female & 0.1557 & 0.2149 & 0.2030\\
    & Male & 0.0802 & 0.2521 & 0.2793\\
    Italy & Female & 0.1527 & 0.2319 & 0.2306\\
    & Male & 0.0567 & 0.1936 & 0.2280\\
    Japan & Female & 0.2302 & 0.4518 & 0.4611\\
    & Male & 0.1880 & 0.5239 & 0.5424\\
    Netherlands & Female & 0.1267 & 0.1661 & 0.1484\\
    & Male & 0.0385 & 0.0457 & 0.0511\\
    New Zealand & Female & 0.0927 & 0.1998 & 0.2015\\
    & Male & 0.0065 & 0.0733 & 0.1029\\
    Norway & Female & 0.0877 & 0.1048 & 0.0889\\
    & Male & 0.0180 & 0.0315 & 0.0360\\
    Sweden & Female & 0.0973 & 0.1874 & 0.1887\\
    & Male & 0.0302 & 0.0312 & 0.0396\\
    Switzerland & Female & 0.1947 & 0.2940 & 0.2657\\
    & Male & 0.1166 & 0.2391 & 0.2404\\
    \hline
  \end{tabular}
\end{table}

\clearpage

\begin{table}[p]
\centering
\caption{Cohorts included in the analysis. The table gives the first and last birth cohorts and the number of cohorts for each population. Both sexes share the same range, yielding twenty-four population--sex series.}
\label{tab:cohorts}
\begin{tabular}{lccc}
\hline
Population & First cohort & Last cohort & Cohorts \\
\hline
Australia               & 1845 & 1910 & 65 \\
Canada                  & 1840 & 1910 & 70 \\
Denmark                 & 1840 & 1910 & 70 \\
England and Wales       & 1840 & 1910 & 70 \\
France                  & 1840 & 1910 & 70 \\
Italy                   & 1840 & 1910 & 70 \\
Japan                   & 1868 & 1910 & 42 \\
Netherlands             & 1840 & 1910 & 70 \\
New Zealand (Non-M\=aori) & 1840 & 1910 & 70 \\
Norway                  & 1840 & 1910 & 70 \\
Sweden                  & 1840 & 1910 & 70 \\
Switzerland             & 1840 & 1910 & 70 \\
\hline
\end{tabular}
\end{table}

\clearpage 


\end{document}